\documentclass[aps,pra,showpacs,twoside,twocolumn,10pt]{revtex4-2}
\usepackage[colorlinks=true, citecolor=red, urlcolor=blue ]{hyperref}
\usepackage{epsfig,newlfont,amssymb,amsfonts,amsmath,bm,subfigure,palatino,mathtools,amsthm,braket,soul,enumitem,color,graphics,graphicx,times,physics,bbold}
\usepackage[normalem]{ulem}
\usepackage{xcolor}
\usepackage{physics}
\usepackage{dsfont}
\usepackage{mathrsfs}
\usepackage{verbatim}
\usepackage{amsthm,amssymb}
\usepackage{comment}
\usepackage{bigints}

\usepackage{gensymb}

\begin{document}


\title{Quantum Heat Transformers}

\author{Arghya Maity$^{1,2}$}

\author{Paranjoy Chaki$^{1}$}

\author{Ahana Ghoshal$^{1,3}$}

\author{Ujjwal Sen$^{1}$}

\address{$^1$Harish-Chandra Research Institute, A CI of Homi Bhabha National
Institute, Chhatnag Road, Jhunsi, Allahabad - 211019, India\\
$^2$School of Physical and Mathematical Sciences, Nanyang Technological University, 21 Nanyang Link, Singapore 637371, Singapore\\
$^3$Naturwissenschaftlich-Technische Fakultät, Universität Siegen, Walter-Flex-Straße 3, 57068 Siegen, Germany}

\begin{abstract} 
 We propose a quantum heat transformer (QHT), a quantum thermodynamic device that modulates temperature gradients between two thermal junctions in quantum systems. Functionally, the QHT is analogous to classical absorption heat transformers in its ability to redistribute thermal energy without external work input. Moreover, we show that its performance ratio mirrors that of classical voltage transformers, where the intrinsic parameters of the system play a role similar to the coil turn ratios.
We initially design the device for a three-qubit system, representing the smallest possible self-contained heat transformer model. Subsequently we extend to four-qubit systems, with a specific emphasis on exploring the step-down mode as the primary focus.  
We showcase the versatility and adaptability of the models by illustrating that a variety of self-contained setups can be constructed, each corresponding to different configurations of the interaction Hamiltonian and their associated self-contained conditions.
An important effect in this study is the proof of existence of a necessarily transient step-down quantum heat transformer, 
that has a dual-mode characteristic, wherein the desired step-down mode can be realized within the transient regime of an originally designed step-up mode of the QHT. We also investigate how to control this transient domain up to which the necessarily transient mode can be achieved, by regulating the initial temperature of the qubits in the four-qubit settings. Therefore, this quantum heat transformer model not only acts as an analog to the classical transformers, but also enjoys advanced characteristics, enabling it to function in both step-up and step-down modes within the same setup, unattainable for classical transformers.
\end{abstract}

\maketitle

\section{Introduction}
\label{sec:intro}
In today’s era of rapid technological progress, quantum devices have become an important milestone in the development of modern technology. These groundbreaking devices harness the principles of quantum mechanics to unlock unprecedented capabilities, promising to reshape industries and revolutionize everyday life.
Unlike traditional electronic devices governed by classical physics, quantum devices operate on the principles of quantum mechanics, where particles can exist in superposition of multiple states and become entangled with one another. This unique behavior enables quantum devices to perform computations, communications, and sensing tasks with efficiency and speed qualitatively higher than their classical counterparts. Examples of such quantum devices include quantum batteries~\cite{q_battery,campaioli2018,Bhattacharjee2021}, quantum heat engines~\cite{Feldmann,Levy,Mitchison2019}, quantum refrigerators~\cite{Palao,Feldmann,Popescu,Levy1,Levy,Correa2014,Mitchison2019}, quantum diodes~\cite{Yuan}, quantum switches~\cite{switch,switch2,switch3}, quantum thermal transistors~\cite{Joulain,Zhang,Su,Mandarino}, quantum sensors~\cite{Sensing_4,Sensing_3,sensing_2,Aslam2023}, quantum thermometers~\cite{Martin-Martinez_2013,Sabin2014,Hofer_2017}, etc.
Among these, quantum refrigerators, quantum heat engines, quantum thermal transistors, and quantum thermometers emerge as pivotal examples of quantum thermal devices, each promising to revolutionize thermal management at the smallest scales. 

In this work, we aim to investigate the potential existence of a new class of quantum thermal devices known as quantum heat transformers (QHTs), designed to manage heat flow in quantum circuits. If such devices are feasible, we further aim to explore their properties. The motivation for this study stems from the fact that heat management plays a crucial role in quantum circuits, particularly in mitigating decoherence and preserving quantum coherence over longer timescales.
By systematically studying the principles, design, and potential applications of quantum heat transformers, we aim to demonstrate their potential significance as practical tools for thermal control in quantum information processing and quantum technologies.

In this investigation, two concepts naturally come to mind: the classical heat transformer and the electrical voltage transformer. These analogies arise because both devices serve the fundamental purpose of transferring energy, the first one in the form of heat and the second in the form of electrical potential energy,
between different parts of a system.
Drawing parallels with these established devices provides a conceptual foundation for envisioning how a quantum heat transformer might operate and what underlying principles it could harness.

A classical heat transformer can upgrade low-temperature heat to a higher temperature without external work input, by using a thermodynamic cycle that redistributes thermal energy. It does not generate additional heat but rather redistributes the supplied thermal energy, delivering a portion at a higher temperature while releasing the remainder at a lower temperature.  A well-known example of a classical heat transformer is the absorption heat transformer (AHT)~\cite{Cudok_Conf_2017, Cudok_RSER_2021}. An AHT is a self-sustained, heat-driven system that upgrades heat to a higher temperature without mechanical work by leveraging the absorption cycle~\cite{Abrahamsson_HRS_1995}. Absorption heat transformers are particularly suitable for heat recovery from industrial processes, with their main advantage being the ability to upgrade waste heat streams to a usable temperature using only negligible amounts of electrical energy and no additional primary energy~\cite{Ma_ATE_2003, Horuz_RE_2010, Liu_JTS_2021}. For these reasons, heat transformers have been described as a ``future technology that will be important for energy utilization in the 21st century” by the International Energy Agency~\cite{Adnan_RE_2007}.

On the other hand, an electrical transformer is a crucial device in power distribution, designed to transfer electrical energy efficiently between circuits through electromagnetic induction \cite{Griffiths,Fitzgerald_Book_2003,purcell}. The fundamental purpose of an electric transformer is to regulate voltage levels for efficient electricity transmission.
When an alternating current (AC) flows through the primary coil, it produces a changing magnetic field in the transformer's core. This changing magnetic field induces a voltage in the secondary coil through electromagnetic induction. The resulting voltage in the secondary coil is proportional to the ratio of number of turns in secondary coil and primary coil, enabling the transformer to step up or step down the input voltage as needed for efficient power distribution.
An electrical transformer regulates the output voltage while maintaining constant power. Transformers are widely used in electronic devices for voltage regulation, impedance matching, and power distribution. They play a vital role in adapting electrical energy to meet the specific requirements of various components within electronic circuits. 
In the realm of electronics, both step-up and step-down transformers are indispensable components, each serving critical functions tailored to specific applications. Nevertheless, step-down transformers find more frequent application and widespread use in various electronic circuits and everyday electronic devices, because when the voltage in the secondary coil is lower than that in the primary coil, it serves to reduce the input voltage to a more manageable level for electronic components.

In this paper, we propose a protocol for a quantum heat transformer, which bears some functional analogies to both the classical heat transformer and the electrical voltage transformer. Unlike electrical transformers that regulate voltage, a QHT operates through temperature adjustment between two terminals. Hence, functionally, QHTs are more analogous to classical heat transformers.
We first design a three-qubit QHT model, representing the smallest possible quantum heat transformer, and subsequently extend our investigation to four-qubit quantum heat transformers. These QHT models are self-contained, meaning they can operate without the need for any external energy sources. Similar to classical voltage
transformers, QHTs comprise two thermal junctions, each consisting of two qubits. The temperature difference between these qubits is the focus of our study. We label these two thermal junctions as the ``primary thermal junction" and ``secondary thermal junction", akin to the primary and secondary coils of a classical transformer, respectively.  In our QHT protocol, we initially assign a temperature to each qubit and subsequently, through interaction, analogous to electromagnetic induction, each qubit undergoes evolution, leading to the generation of new temperature differences at both the primary and secondary junctions at each evolved time step. In classical transformer the voltage difference that a transformer can increase or decrease in the secondary coil is determined by its turns ratio.
In the QHT protocol, the nature of the interaction and the initial assigned temperature of the qubits allows us not only to achieve step-up or step-down transformations but also to finely regulate the 
performance ratio ($\mathcal{R}_T$), defined as the ratio between the temperature gradients of the secondary and primary junctions. Specifically, our investigation centers on the step-down quantum heat transformer configuration, where a step-down transformer is defined as one with a lower temperature gradient at the secondary thermal junction compared to the primary one.
Note that although the quantum heat transformer is structurally analogous to a classical voltage transformer and operates in a similar way but by considering heat instead of voltage, and exhibits certain behaviors similar to classical voltage transformers, it differs in a fundamental way. In contrast to a classical voltage transformer, which ideally conserves power in a perfectly efficient manner, the quantum heat transformer inevitably suffers energy losses due to decoherence afflicted by the heat baths to which it is connected by construction, making it difficult to formulate a direct analog of energy (or power) conservation in the quantum regime. However, since it can operate in both step-up and step-down modes, it can still be regarded as a genuine transformer rather than merely a dissipative element like a thermal resistor or rheostat.

The rest of the paper is arranged as follows. In Sec.~\ref{Sec:2}, we introduce the concept of a QHT and provide a general discussion of the device. We also outline its functional analogy with classical heat transformers and define the corresponding performance measure in this section. In Sec.~\ref{Sec:2a}, we discuss the three-qubit QHT models and the associated self-contained interaction Hamiltonians. In this section, we also elucidate the operational scheme of the QHTs and establish a relation between the ratio of temperature gradients and specific system parameters, which exhibits a functional resemblance to the voltage-to-coil-turn ratio in a classical voltage transformer.
In Sec.~\ref{Sec:3}, we explore the existence of necessarily transient step-down quantum heat transformers, which operates as a step-up transformer in the steady-state regime, but operate as a step-down transformer at some fixed transient times. We present the operational characteristics of a self-contained four-qubit QHT in Sec.~\ref{Sec:4}. Lastly, in Sec.~\ref{Sec:con}, we present the concluding remarks. 

\section{Quantum heat transformers}
\label{Sec:2}

A quantum heat transformer is a quantum thermodynamic device designed to control and manipulate heat flows at the microscopic level using quantum systems such as qubits.  The primary aim of a QHT is to modulate the temperature gradient between two thermal junctions by redistributing thermal energy in a controlled manner.

In a typical QHT setup with qubits, each qubit is coupled to a thermal bath maintained at a specific temperature, and the overall dynamics of the QHT are governed by the principles of open quantum system. These qubits are divided into two groups, which define the primary and secondary thermal junctions, respectively. The definition of the temperature gradient for each junction is model-dependent and may vary depending on how the junctions are constructed and how temperatures are assigned. The goal of the QHT is to manipulate the temperature gradient in the secondary junction in relation to that in the primary junction, thereby achieving a thermal transformation.


While the primary objective of this work is to model a quantum analogue of the classical transformers, our initial focus is on developing a self-contained minimal qubit model for QHT. By the term ``self-contained", we refer to systems that require no external work input or external control to operate. Instead, the functioning of the quantum heat transformer relies solely on the internal interactions among the qubits and their coupling to the thermal baths. This notion of self-containment is consistent with the widely accepted framework introduced by Linden, Popescu, and Skrzypczyk in their seminal work on the self-contained quantum refrigerator~\cite{Popescu}, where the system autonomously extracts heat from a cold bath without external driving or control. Also, in Ref.~\cite{Correa2014}, it is shown that heat exchange among three thermal reservoirs enables autonomous cooling without any external work input. Moreover, it is also demonstrated that the performance of the absorption refrigerator can be further enhanced through quantum reservoir engineering.
 Similarly, our QHT model operates autonomously, governed entirely by internal interactions and thermal contact, and performs thermal energy redistribution without the need for external work sources. Self-contained quantum devices hold significant importance in various fields of research and practical applications~\cite{Popescu,Skrzypczyk_2011}. This autonomy enhances their reliability and robustness, particularly in scenarios where access to external power sources is limited or impractical. Here, we intend to begin with a self-contained three-qubit protocol, where one qubit serves as a common junction for both the primary and secondary thermal junctions. Subsequently, we will explore a four-qubit protocol, wherein the architecture dictates the absence of any shared qubit between the primary and secondary junctions of the transformer. As we already mentioned, both types of transformers play crucial roles in various contexts, however, step-down transformers prove to be particularly pertinent to routine activities. Consequently, we have chosen to prioritize the step-down configuration as the primary focus for studying different models of QHT. Nonetheless, we also present specific cases demonstrating the step-up operation of QHTs. 


\subsection{Performance Measure for QHTs}
Based on how the temperature gradient in the secondary junction responds to the primary junction, the QHT can operate in two distinct regimes. If the temperature gradient in the secondary junction becomes larger than that in the primary, the device is said to operate in a step-up mode. Conversely, if the gradient in the secondary junction becomes smaller, the device is functioning in a step-down mode. To quantitatively assess this behavior, we define a measure of performance known as the performance ratio, denoted by $\mathcal{R}_T$. This is defined as the ratio between the temperature gradient in the secondary junction and that in the primary junction
\begin{equation}
    \mathcal{R}_T=\frac{\Delta T_s}{\Delta T_p}.
\end{equation}
Here, $\Delta T_p$  and $\Delta T_s$ refer to the temperature gradients across the primary and secondary junctions, respectively. The ratio $\mathcal{R}_T$ provides a clear classification of the operational regime of the QHT. Specifically, the transformer operates in a step-down mode when $\mathcal{R}_T < 1$, and in a step-up mode when $\mathcal{R}_T > 1$. 

The performance ratio $\mathcal{R}_T$ in our quantum heat transformer can be analogized to the voltage ratio $\mathcal{R}_v=V_s/V_p$, where $V_s$ and $V_p$ are the voltage differences across the secondary and primary coils, respectively, in a classical voltage transformer.  
We will discuss this analogy in more detail in Sec.~\ref{Sec:PR}.

\subsection{Functional analogy with a classical heat transformer}
\begin{figure}
\fbox{\includegraphics[width=7.0cm, height=4.8cm]{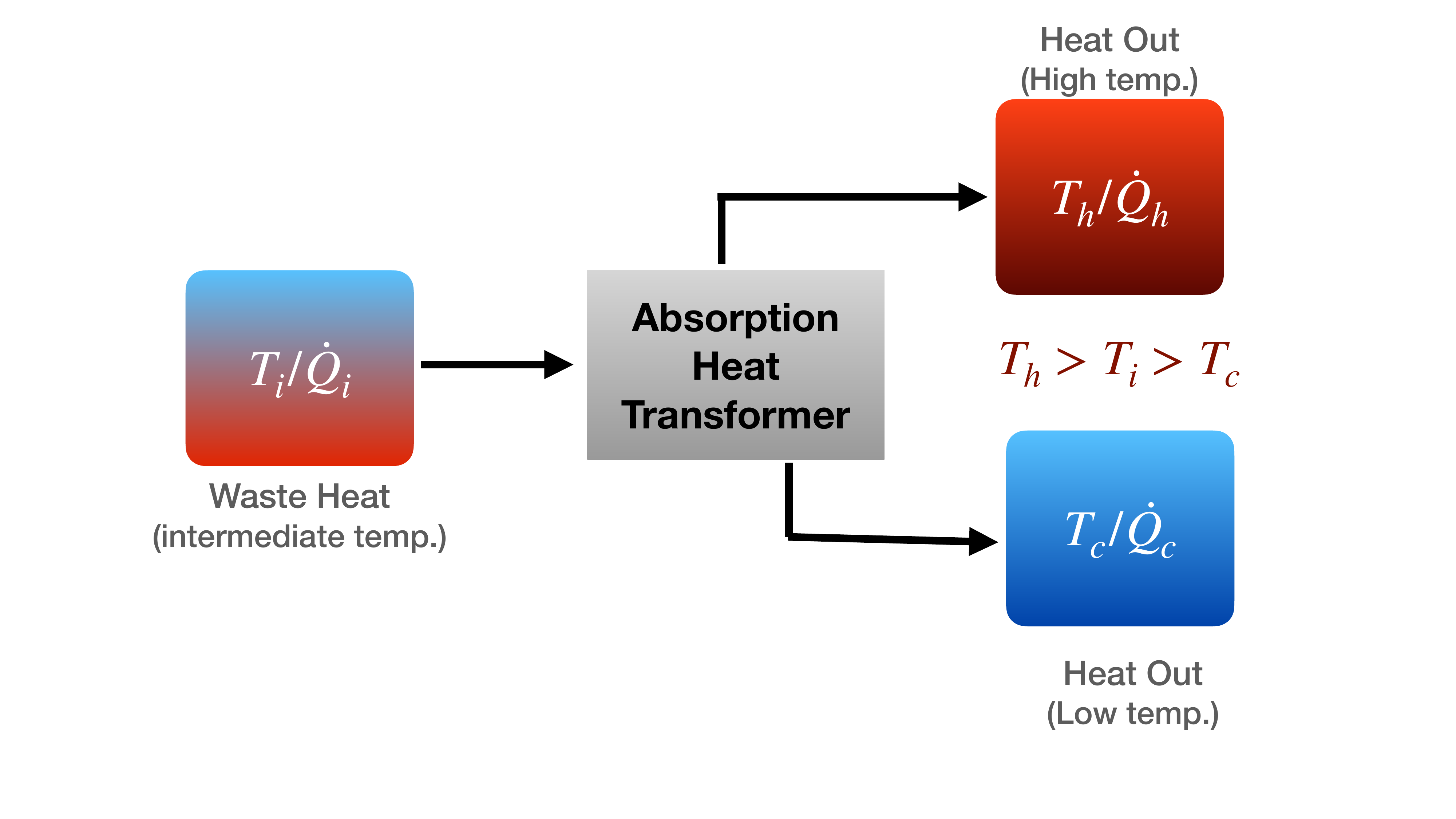}}
\caption{Schematic diagram showing the operation of an absorption heat transformer. An Absorption Heat Transformer takes waste heat (which is usually discarded) and converts part of it to a higher temperature, making it usable again—without using much electricity. It works on a thermodynamic cycle involving absorption and desorption processes. In the diagram waste heat at intermediate temperature ($T_i$) is partially upgraded to higher temperature ($T_h$) output, while rejecting the remaining heat at a lower temperature ($T_c$) , satisfying $T_h > T_i > T_c$.
}
\label{Fig:AHT}
\end{figure}
A quantum heat transformer can be viewed as the quantum counterpart of a classical heat transformer, particularly resembling the functionality of an absorption heat transformer~\cite{Cudok_RSER_2021, Cudok_Conf_2017}. 
It is driven by a heat flow $\dot{Q}_{i}$ at an intermediate temperature level $T_{i}$, which is split into an upgraded heat flow $\dot{Q}_{h}$ at a higher temperature $T_{h}$ and a downgraded heat flow $\dot{Q}_{c}$ at a lower temperature $T_{c}$ (see Fig.~\ref{Fig:AHT}).
The efficiency of an AHT is measured by the coefficient of performance, defined as the ratio of the upgraded heat flow $\dot{Q}_{h}$ to the driving heat flow $\dot{Q}_{i}$: $\text{COP} = \dot{Q}_{h} / \dot{Q}_{i}$.
Typical COP values range from $0.3$ to $0.5$, meaning that $30–50\%$ of the input heat is upgraded.
The performance of a heat transformer decreases with increasing temperature lift ($T_h - T_i$) and improves with greater temperature thrust ($T_i - T_c$). 
Like classical heat transformers, a QHT redistributes thermal energy, enabling heat to be transferred from lower to higher temperatures without external work input. Hence, the quantum heat transformer model serves as a quantum analog to the classical heat transformer. Just as classical heat transformers upgrade thermal energy using thermodynamic cycles, a QHT 
manipulates heat flow at the microscopic level. 

We now turn to specific quantum heat transformer models. In constructing quantum analogues of classical devices, researchers aim to use the smallest possible system—typically the minimum number of qubits required to perform the task. For instance, a three-qubit setup can realize a self-contained quantum refrigerator, while a single qubit can serve as a reservoir in a quantum heat engine. In our QHT model, we show that three qubits are sufficient to implement the transformer protocol, and we further explore an extended version using four qubits.


\section{Self-contained three-qubit QHT model}
\label{Sec:2a}


We begin by considering three qubits, labeled as $q_1$, $q_2$, and $q_3$, each locally connected to three thermal reservoirs at temperatures $\tilde{\tau}_1$, $\tilde{\tau}_2$, and $\tilde{\tau}_3$, respectively, and each qubit is in thermal equilibrium with its associated reservoir. Let $q_2$ denote the qubit serving as the common qubit for both the primary and secondary thermal junctions. 
We take the free Hamiltonian of the three-qubit composite system as $\mathcal{H}_{0} = \frac{\mathcal{K}}{2} \sum_{j=1}^{3} \mathcal{E}_{j}\sigma_{j}^{z}$, where $\sigma^{z}_j$ is the $z$ component of the Pauli matrix and $ \mathcal{K} \mathcal{E}_j$ is the energy gap between the two levels of the $j^{\text{th}}$ qubit. Here, $\mathcal{K}$ is a constant with the unit of energy and $\mathcal{E}_j$s are dimensionless energy parameters. 
Now, the crucial task is to select the interaction Hamiltonian for this three-qubit system that will effectively demonstrate the properties of QHT while ensuring the protocol remains self-contained. To meet the self-contained condition, we find that, for the free system Hamiltonian $\mathcal{H}_0$, various interaction Hamiltonians can be crafted. Without assuming any of the $\mathcal{E}_j$ to be zero, the three-qubit free Hamiltonian is subject to some specific self-contained conditions, as follows:
\begin{align}
\label{Equ:canonical}
\mathcal{H}_{\text{int}}^{1} = & \mathcal{K} g(|111\rangle \langle 000| + h.c); \nonumber ~~ \mathcal{E}_3=-\mathcal{E}_2-\mathcal{E}_1, \\
 \mathcal{H}_{\text{int}}^{2} = & \mathcal{K} g(|101\rangle \langle 010| + h.c); \nonumber ~~ \mathcal{E}_3=\mathcal{E}_2-\mathcal{E}_1, \\
 \mathcal{H}_{\text{int}}^{3} = &  \mathcal{K} g(|110\rangle \langle 001| + h.c); \nonumber ~~ \mathcal{E}_3=\mathcal{E}_2+\mathcal{E}_1, \\
 \mathcal{H}_{\text{int}}^{4} = & \mathcal{K} g (|100\rangle \langle 011| + h.c); ~~ \mathcal{E}_3=-\mathcal{E}_2+\mathcal{E}_1.
\end{align}
Here, $\ket{1}$ and $\ket{0}$ are the eigenvectors of $\sigma_z$ matrix corresponding to the eigenvalues $-1$ and $+1$, respectively.
$g$ is the dimensionless coupling strength of the interaction. 
As each qubit resides in its respective thermal state initially, the state of each qubit at the initial time, $\tilde{t}=0$, can be expressed as
\begin{equation}
\rho_j(0)=p_j^{\text{in}}\ket{0}\bra{0}+(1-p_j^{\text{in}})\ket{1}\bra{1},
\label{eq:state}
\end{equation}
for $j=1$, $2$, and $3$ and the composite three-qubit initial state is given by $\rho(0)=\rho_1(0)\otimes \rho_2(0)\otimes \rho_3(0)$. Here, $p_j^{\text{in}}$ is the initial probability to find the qubit in the state $\ket{0}$. 
Now, considering the coupling between the systems and their respective reservoirs as weak, we are able to make use of the Born-Markov and secular approximations~\cite{Sudarshan_JMP_1976,Lindblad_CMP_1976,Petruccione_book,Alicki_2007,Rivas_Huelga_book,Subhashish_Banerjee_book,Lidar_2020_lecture} to govern the evolution of the composite system under the influence of the local thermal baths. Consequently, the dynamics of the reduced system, $\rho(t)$, can be described using the Gorini-Kossakowski-Sudarshan-Lindblad (GKSL) master equation, as
\begin{equation}
    \frac{d \rho(t)}{d t}= -i [ \mathscr{H}_{0}+\mathscr{H}_{\text{int}}, \rho(t)] + \frac{\hbar}{\mathcal{K} } \mathcal{L}(\rho(t)).
    \label{Equ:Master_Equ}
\end{equation}
Here, we introduce the dimensionless time as $t=\mathcal{K} \Tilde{t}/\hbar$. We define $\mathscr{H}_{0} = \mathcal{H}_{0}/\mathcal{K}  $, and $\mathscr{H}_{\text{int}} = \mathcal{H}_{\text{int}}^x/\mathcal{K} $, where $x$ can be any of $1$, $2$, $3$, and $4$.  The dissipative term, $\mathcal{L}(\rho(t)) =\sum_j \mathcal{L}_j(\rho(t))=\sum_{j,\omega} \gamma_{j}(\omega) \left [ A_{j}(\omega) \rho(t) A_{j}^{\dagger}(\omega) - \frac{1}{2} \{ A_{j}^{\dagger}(\omega) A_j(\omega), \rho(t)  \}
 \right] $, arises due to the presence of thermal environments. Here $j$ runs from $1$ to $3$. Additionally, $\omega$ represents the transition frequencies, $\gamma_j(\omega)$ denotes the decay constants, and $A_j(\omega)$ are the Lindblad operators. Suppose $\lambda_1$ and $\lambda_2$ represent two eigenvalues of $\mathcal{H}_0+\mathcal{H}^x_{\text{int}}$. Then, the transition frequency linked to the transition between these two energy eigenstates is given by $\omega=\frac{1}{\hbar}(\lambda_2-\lambda_1)$.  See~\cite{Sudarshan_JMP_1976,Lindblad_CMP_1976,Petruccione_book,Alicki_2007,Rivas_Huelga_book,Lidar_2020_lecture} for detailed discussions on transition energies, decay constants, and the construction of Lindblad operators. For the validation of Born-Markov and secular approximations, the condition, $\max\{\frac{\hbar}{\mathcal{K}}\gamma_j\}\ll \min\{|\mathcal{E}_j|,g\}$, has to be satisfied. We have verified numerically that this condition holds for all parameter regimes considered in this work. 
 Since the free Hamiltonian and the interaction Hamiltonians satisfy $[\mathcal{H}_0,\mathcal{H}_{\text{int}}^x]=0$ for $x\in \{1,2,3,4\}$, the field energies and the interaction energy are independently conserved under closed-system evolution. The initial three-qubit state $\rho(0)$ is diagonal in the eigenbasis of $\mathcal{H}_0$, and the only off-diagonal elements can be generated by the interaction Hamiltonian $H_{\text{int}}^x$. Furthermore, the open-system dynamics is governed by a GKSL master equation, whose dissipator cannot create coherence in this basis. As a result, the reduced state of each qubit remains diagonal at all times with respect to the eigenbasis of $\mathcal{H}_0$~\cite{Das_2019}.
Hence, the reduced density matrices of the individual subsystems, denoted as $\rho_j(t)$ at any given time $t$, retain their diagonal form as given in Eq.~(\ref{eq:state}). However, in this evolving scenario, the equilibrium probabilities $p_j^{\text{in}}$ are replaced by time-dependent probabilities $p_j(t)$ and it is connected to the local temperature of each qubit at time $t$ as  
$p_j(t)=\frac{1}{1+\exp(\mathcal{K}\mathcal{E}_j/k_B\tilde{T}_j(t))}$, where $k_B$ is the Boltzmann constant and the index $j$ goes to $1$, $2$, and $3$. 
From now on, we define the dimensionless temperatures of the qubits as 
$T_j(t)= \frac{k_{B}}{\mathcal{K} } \Tilde{T}_j(t)$ and those of the baths as $\tau_j=\frac{k_{B}}{\mathcal{K} } \Tilde{\tau}_j$ for $j=1$, $2$, and $3$. Thus, at every instant of time, we have the temperature gradients, $\Delta T_p(t)=|T_1(t)-T_2(t)|$ for the primary junction, and $\Delta T_s(t)=|T_3(t)-T_2(t)|$ for the secondary junction. Note that, here we define $\Delta T_p$ and $\Delta T_s$ as the absolute values of the temperature differences across the primary and secondary thermal junctions, respectively. By taking absolute values, we quantify only the magnitude of each thermal bias, not its direction. This convention allows us to define the performance ratio $\mathcal{R}_T= \Delta T_{s}/ \Delta T_{p}$ in terms of the relative strength of the two gradients—i.e., how large the secondary bias is compared to the primary—without regard to the actual direction of heat flow. 

\begin{figure}
		\centering
	\includegraphics[scale=0.2]{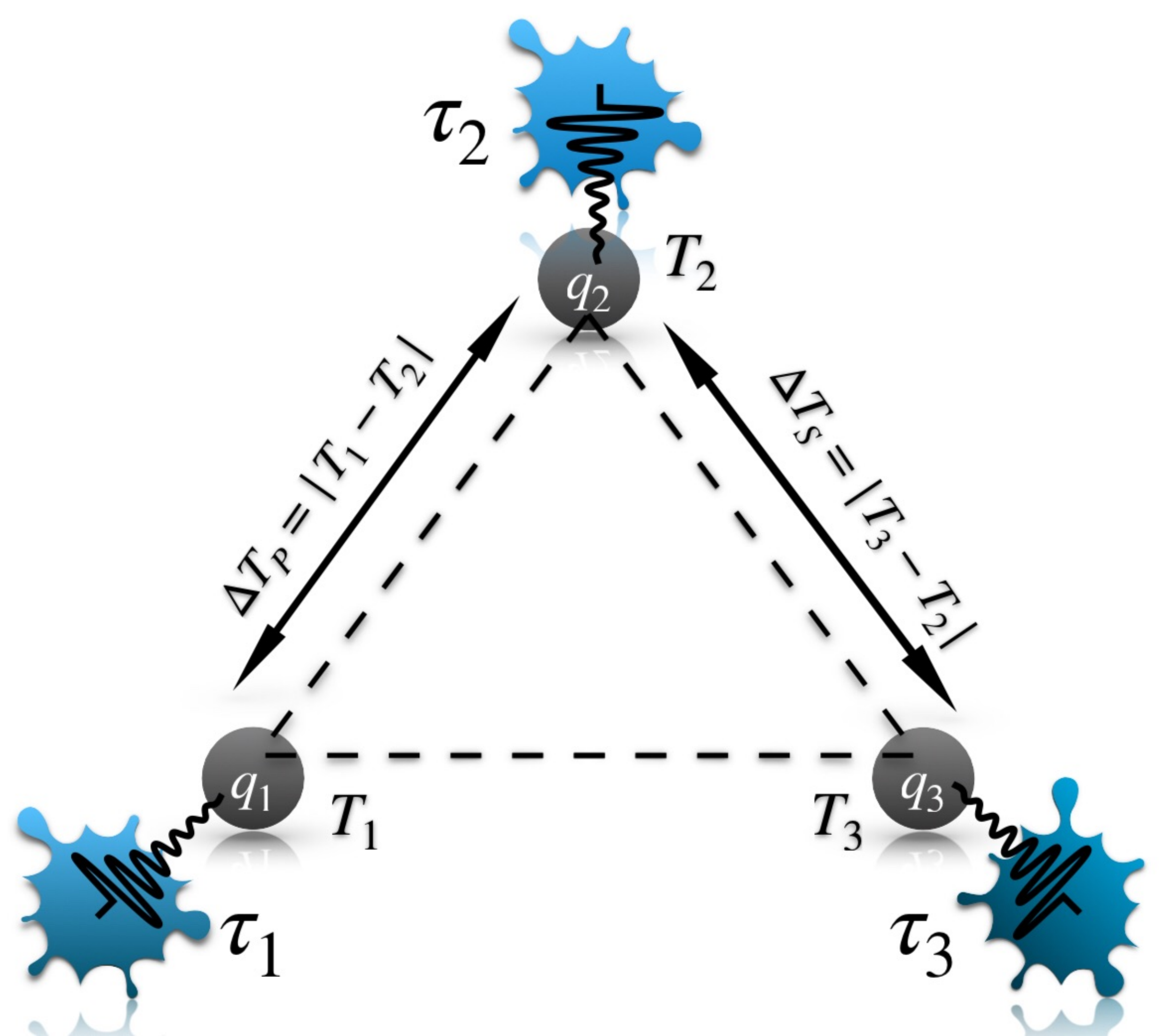}
		\caption{Schematic diagram of a three-qubit QHT. Here, $q_1$, $q_2$, and $q_3$ represent the three qubits of the quantum transformer, each immersed in a separate thermal bath at dimensionless temperatures $\tau_1$, $\tau_2$, and $\tau_3$, respectively. The pair of qubits $q_1$ and $q_2$ forms the primary thermal junction, while the pair $q_2$ and $q_3$ constitutes the secondary thermal junction. At any given time $t$, the temperatures of these qubits are represented as $T_1(t)$, $T_2(t)$, and $T_3(t)$, where  $\tau_j=\frac{k_{B}}{\mathcal{K} } \Tilde{\tau}_j$ and $T_j(t)= \frac{k_{B}}{\mathcal{K} } \Tilde{T}_j(t)$. The temperature gradient across the primary thermal junction is defined as $\Delta T_p(t) = |T_1(t) - T_2(t)|$, while across the secondary junction it is denoted as $\Delta T_s(t) = |T_3(t) - T_2(t)|$.}
		\label{fig:3qubit-diagram}
	\end{figure}

A schematic diagram of this three-qubit QHT is depicted in Fig.~\ref{fig:3qubit-diagram}. To design a step-down quantum transformer, it is essential to ensure that, at any time $t$, $\Delta T_p(t) > \Delta T_s(t) $.

We now analyze the four self-contained conditions for designing the three-qubit QHTs, given in Eq.~(\ref{Equ:canonical}). Let us consider the initial temperature of the three qubits, i.e., the temperature of the three reservoirs follows the condition $\tau_1 > \tau_2 > \tau_3$, and also $\mathcal{E}_2>\mathcal{E}_1 > 0$. In this scenario, one approach to designing an effective QHT involves an increase in the temperature of $q_1$ over time, accompanied by a corresponding decrease in the temperature of $q_2$. Consequently, the temperature difference $\Delta T_{p}(t) = |T_{1}(t) - T_{2}(t)|$ at the primary junction at time $t$ is expected to increase with time.
Conversely, as the temperature of $q_2$ decreases, the temperature of $q_3$ should increase with time. This adjustment ensures that the temperature difference at the secondary junction $\Delta T_{s}(t) = |T_{3}(t) - T_{2}(t)|$ decreases over time. Therefore, under these circumstances, the condition demands $\mathcal{H}_{\text{int}}^{2}$ as the interaction Hamiltonian between the qubits, with the choices of parameters facilitating the system transitioning from an initial state of $|101\rangle$ to the state $|010\rangle$ as time progresses with more probability than the opposite process. Note that the interaction Hamiltonian $\mathcal{H}^2_{\text{int}}$ is commonly utilized in the design of quantum refrigerators, as discussed in the literature~\cite{Popescu}. However, in this QHT protocol involving the interaction Hamiltonian $\mathcal{H}_{\text{int}}^{2}$, does not exhibit the characteristics of a quantum refrigerator. 
The objective of refrigeration is typically achieved when the qubit with the lowest initial temperature among the three qubits undergoes cooling. Yet, in our protocol, the qubit having the lowest temperature, undergoes a transition to a higher excited state, indicating an increase in its temperature over time. This difference illustrates that the same Hamiltonian structure can support different functionalities depending on the operational goal and parameter regime.
Now, if we alter the considerations and designate $q_3$ as the common qubit, with the initial temperatures of the three reservoirs following the sequence $\tau_1 > \tau_3 > \tau_2$, and $\mathcal{E}_1>0$, $\mathcal{E}_2>0$, a parallel methodology reveals $\mathcal{H}_{\text{int}}^{3}$ as the compatible interaction Hamiltonian. By appropriately selecting parameters with this interaction Hamiltonian, we can ensure that the system initially resides in the $|110\rangle$ state with a higher probability and transitions to the $|001\rangle$ state as time progresses.  This transition ensures that over time, the temperature of $q_1$ increases while correspondingly, the temperature of $q_3$ decreases, thereby causing the primary junction temperature difference  $\Delta T_{p}(t) = |T_{1}(t) - T_{3}(t)| $ to increase. Also, the temperature of $q_2$ rises over time to decrease the secondary junction temperature difference $\Delta T_{s}(t) = |T_{2}(t) - T_{3}(t)| $. 
Similarly, if we designate $q_1$ as the common qubit, $\mathcal{H}_{\text{int}}^4$ emerges as the most appropriate Hamiltonian for constructing a QHT for some fixed set of initial conditions. 
In this way, we can formulate diverse protocols for the QHT model tailored to different initial conditions.

\begin{figure*}
\includegraphics[width=5.9cm, height=4.5cm]{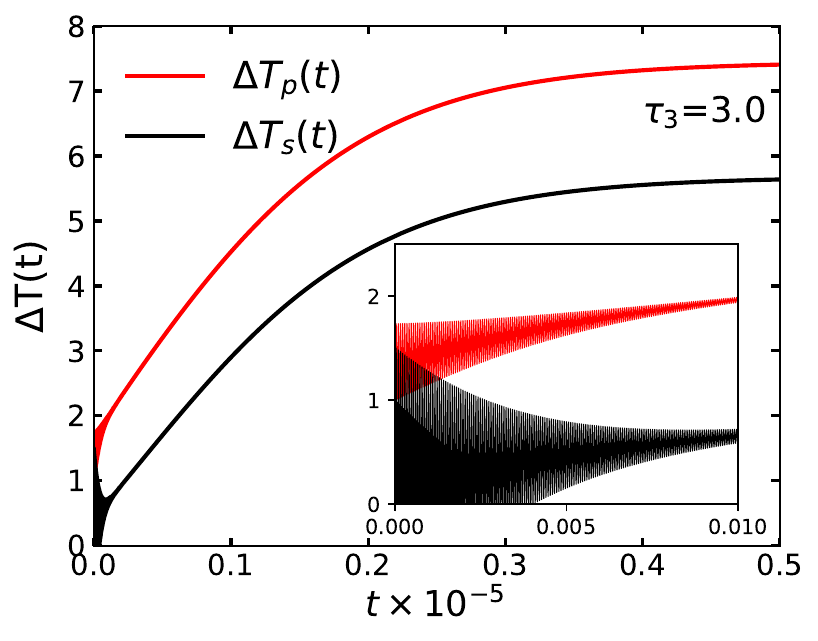}
\includegraphics[width=5.9cm, height=4.5cm]{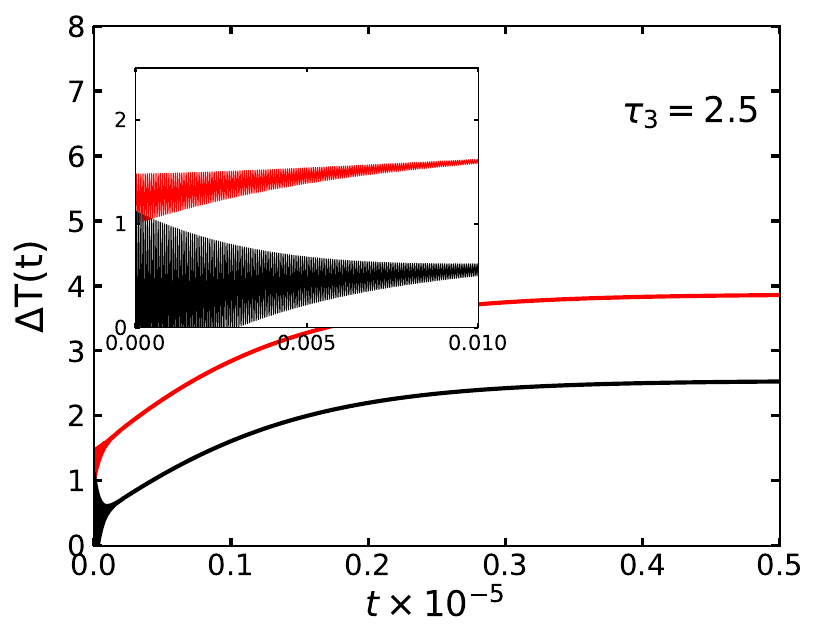}
\includegraphics[width=5.9cm, height=4.5cm]{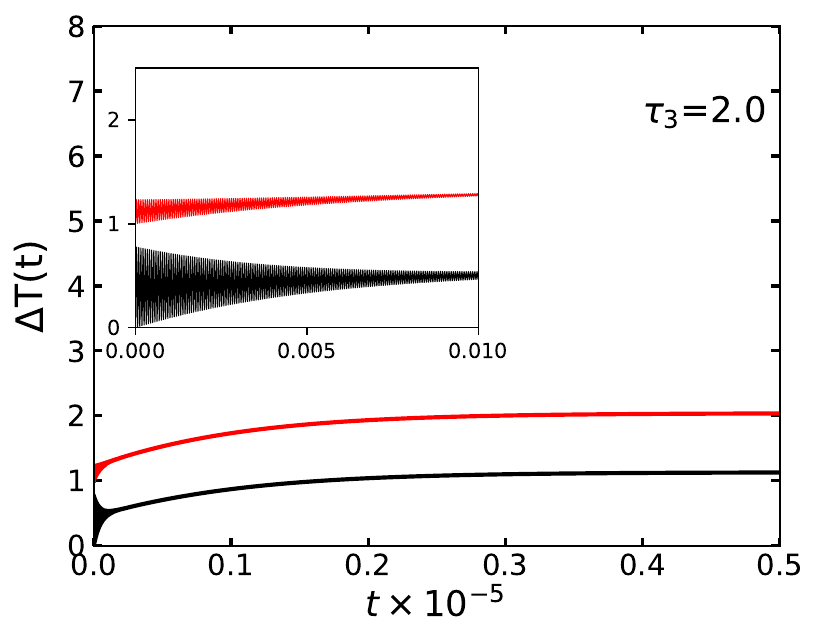}
\caption{Time evolution of temperature gradients at primary and secondary thermal junctions, with the variation in the initial temperature of $q_3$ for the interaction Hamiltonian $\mathcal{H}_{\text{int}}^1$. In each panel, the temperature gradient of the primary thermal junction, $\Delta T_p(t)$, is represented by red lines, while that of the secondary thermal junction, $\Delta T_s(t)$, is denoted by black lines. The insets illustrate the initial behaviors of the temperature gradients. As the red curves lie above the black ones, it indicates that the QHTs are operating as step-down transformers.
Here, we set $\mathcal{E}_1=1.0$, $\mathcal{E}_2=2.0$, $\tau_1=1.0$, and $\tau_2=2.0$. The self-contained condition of $\mathcal{H}_{\text{int}}^{1}$ yields $\mathcal{E}_3 = -\mathcal{E}_2 - \mathcal{E}_1$. Additionally, we choose the interaction strength $g=0.5$. The dimensionless coupling strength parameters of the Ohmic spectral density function are taken to be $\alpha_1 = 10^{-4}$, $\alpha_2 = 10^{-5}$, and $\alpha_3=10^{-3}$, while the cutoff frequencies of the baths are $\Omega_1=\Omega_2=\Omega_3=10^3$. All quantities plotted here are dimensionless.}
\label{Fig:H_int_111-000}
\end{figure*}

\subsection{QHT model : The interaction Hamiltonian $\mathcal{H}_{\text{int}}^{1}$}


For a more deeper insight into the operational characteristics of QHTs, we will now delve into the operational aspects of QHTs utilizing the interaction Hamiltonian $\mathcal{H}_{\text{int}}^1$.
Let us consider a scenario where the three qubits are interacting through $\mathcal{H}_{\text{int}}^1$, and the three thermal baths are bosonic baths, each consisting of an infinite number of harmonic oscillators. Here we set $\mathcal{E}_1>0$ and $\mathcal{E}_2>0$. The dynamics of the three-qubit system are governed by the master equation given in Eq.~(\ref{Equ:Master_Equ}), with all relevant quantities detailed in App.~\ref{appen:1}. For this case, the transition frequencies are $\omega \in \{\mathcal{E}_{1},\,\mathcal{E}_{1}\pm g,\; \mathcal{E}_{2},\,\mathcal{E}_{2}\pm g,\; \mathcal{E}_{3},\,\mathcal{E}_{3}\pm g\}$.

Let us first discuss the attainability of the self-contained condition. If we impose the constraint $\mathcal{E}_3 = -\mathcal{E}_1 - \mathcal{E}_2$, then $\mathcal{H}_0$ possesses two degenerate eigenstates: $\ket{000}$ and $\ket{111}$. Now, if the initial conditions on $\mathcal{E}_1$, $\mathcal{E}_2$ and the temperatures of the baths are such that the probability of being in $\ket{111}$ is higher than that of $\ket{000}$, i.e.,
\begin{align}
\mathcal{P}^{\text{in}}_{111} &> \mathcal{P}^{\text{in}}_{000},
\label{equ_prob_111_000}
 \end{align}
where $\mathcal{P^{\text{in}}}_{111}=(1-p_1^{\text{in}})(1-p_2^{\text{in}})(1-p_3^{\text{in}})$ and $\mathcal{P}^{\text{in}}_{000}=p_1^{\text{in}} p_2^{\text{in}} p_3^{\text{in}}$, then activating the interaction Hamiltonian $\mathcal{H}_{\text{int}}^{1}=\mathcal{K} g \left( \ket{111} \bra{000} + \text{h.c.} \right)$ induces a transition from $\ket{111}$ to $\ket{000}$. Conversely, if $\ket{000}$ is initially more probable, the transition occurs in the opposite direction. Crucially, since both states are degenerate and have the same total internal energy, these transitions occur without any net energy cost, ensuring energy conservation. This allows the system to operate autonomously, without any external work input,  thereby fulfilling the self-contained condition. The third qubit and the specific form of interaction are crucial for establishing the self-contained condition; without them, a two-qubit setup would generally require external work or control to enable energy exchange, thus violating autonomy. See Ref.~\cite{Popescu} for a detailed discussion on how the self-contained condition is established as well as some other significant research work related to it~\cite{Giovannetti_PRL_2013, Huang_PRL_2024}.

Let us now choose the parameters of the system and the interaction Hamiltonians to ensure that the initial probability of finding the qubits in the state $\ket{111}$ exceeds that of $\ket{000}$. 
In this scenario, qubits $q_1$ and $q_2$ are initially in their ground state, while $q_3$ is in the excited state. With the progress of time, qubits $q_1$ and $q_2$ will transition to their excited state, while $q_3$ will transition to the ground state. 
Note that the reverse condition of Eq.~(\ref{equ_prob_111_000}) is also feasible, but we have opted for this particular choice arbitrarily.
In Fig.~\ref{Fig:H_int_111-000}, we illustrate the time evolution of the temperature gradients of the primary ($\Delta T_p(t)$) and secondary ($\Delta T_s(t)$) thermal junctions, by changing the initial temperature of one of the qubits, say the third qubit $q_3$.
Here we set the initial temperature of $q_1$ as $T_1(0)=\tau_1=1.0$ and the initial temperature of $q_2$ as $T_2(0) = \tau_2= 2.0$, and vary the value of $T_3(0)$. 
For the initial investigation, we set the initial temperature of $q_3$ as $T_3(0) = \tau_3 = 3.0$. This ensures that at time $t=0$, we have $\Delta T_{p}^i = \Delta T_{p}(0) = |T_{1}(0) - T_{2}(0)| = 1.0$ and $\Delta T_{s}^{i} = \Delta T_{s}(0) = |T_{3}(0) - T_{2}(0)| = 1.0$. Hence, the difference between the primary and secondary junctions starts at zero. If we now allow the system to evolve under the influence of the thermal baths, we can observe that the difference between $\Delta T_p(t)$ and $\Delta T_s(t)$ increases over time until it saturates at a value of $1.768$. See the first panel of Fig.~\ref{Fig:H_int_111-000}.
Subsequently, as the initial temperature of $q_3$ decreases compared to its previous value in different scenarios, as studied in the subsequent panels of Fig.~\ref{Fig:H_int_111-000}, the differences between the saturated values of $\Delta T_p(t)$ and $\Delta T_s(t)$ reduce, and their differences also decrease.  Thus, by adjusting the initial temperature of $q_3$, we can effectively regulate the performance of the transformer. 

\subsubsection{Analysis of the step-down QHT effect}

If we examine the interaction Hamiltonian $\mathcal{H}_{\text{int}}^1$ and its self-contained condition, it becomes evident that it yields a negative $\mathcal{E}_3$ value, regardless of the chosen values for $\mathcal{E}_1$ and $\mathcal{E}_2$. Consequently, for qubit $q_3$, the state $\ket{0}$ corresponds to the ground state, while the state $\ket{1}$ corresponds to the excited state.
When we examine each qubit individually, we observe that the temperatures of $q_1$ and $q_2$ increase over time $t$. However, the temperature of $q_3$ initially decreases for a short period before stabilizing back near to its initial value. So, this QHT scheme is not as straightforward as the previous two cases discussed for $\mathcal{H}_{\text{int}}^2$ and $\mathcal{H}_{\text{int}}^3$. The most effective approach to comprehend this protocol is by analyzing the rates of temperature change relative to the evolution time, rather than focusing solely on the individual temperature changes of each qubit.
The rationale behind constructing the QHT framework using $\mathcal{H}_{\text{int}}^1$
lies in the fact that the rates of temperature change with respect to evolution time $\left ( \frac{\partial T_j(t)}{\partial t} \right )$ differ among the qubits. 
From Fig.~\ref{Fig:H_int_111-000}, we can infer that the profile of the temperatures is oscillatory, indicating that the quantities $ \frac{\partial T_j(t)}{\partial t}$ will also oscillate around zero, having positive and negative values. Therefore, the rate at which the temperature increases will depend on the absolute values of $ \frac{\partial T_j(t)}{\partial t}$. 

\begin{figure}[t]
\includegraphics[width=8.5cm, height=5.8cm]{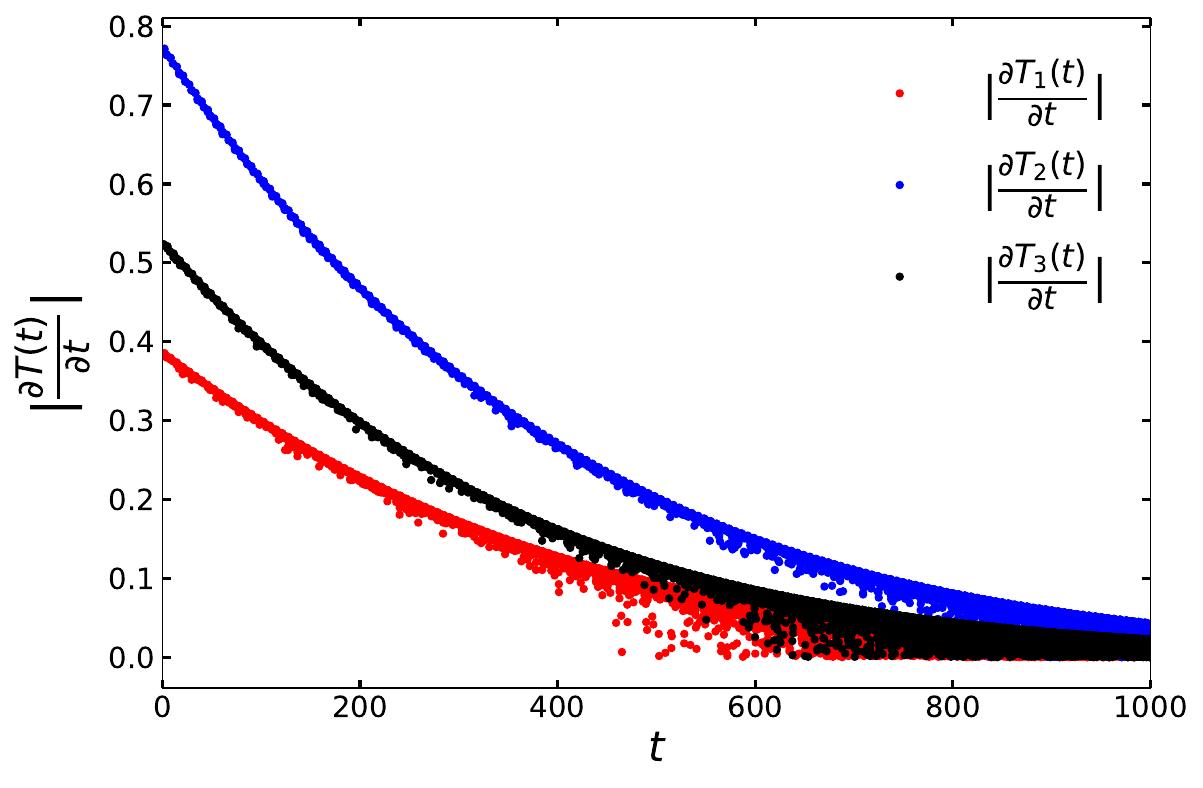}
\caption{Absolute rates of change in temperatures of the three qubits with time. Here we plot the quantities $\Big |\frac{\partial T{j}(t)}{\partial t}\Big |$ for $j=1$, $2$, and $3$ vs. $t$ for the interaction Hamiltonian $\mathcal{H}_{\text{int}}^1$. Here, we set $\tau_3=3.0$.
All other considerations are same as in Fig.~\ref{Fig:H_int_111-000}. 
All the quantities plotted here are dimensionless.}
\label{Fig:rate_111_000}
\end{figure}

Now, to increase the temperature difference between two qubits associated with the primary junction $( \Delta T_{p}(t) ) $ over time through the interaction Hamiltonian $\mathcal{H}_{\text{int}}^{1}$, it is crucial for the absolute rate of temperature increase of $q_1$ $\left (\Big | \frac{\partial T_{1}(t)}{\partial t} \Big |\right )$ to be less than the absolute rate of temperature increase of $q_2$ $\left ( \Big | \frac{\partial T_{2}(t)}{\partial t} \Big | \right )$. Consequently, to reduce the temperature difference between two qubits associated with the secondary junction $( \Delta T_{s}(t) ) $ over time, it is necessary for the rate of temperature increase of $q_3$ $\left ( \Big | \frac{\partial T_{3}(t)}{\partial t} \Big |\right )$ to be less than the rate of temperature increase of $q_2$ $\left ( \Big | \frac{\partial T_{2}(t)}{\partial t} \Big | \right )$.  Additionally, between $q_1$ and $q_3$, the rate of temperature increase of $q_1$ should be less than the rate of temperature increase of $q_3$. So, we can write the order of rate of change of temperature of qubits in QHT model involving the interaction Hamiltonian $\mathcal{H}_{\text{int}}^{1}$ as follows,
\begin{equation}
\label{eq:rates}
    \Big |\frac{\partial T_{1}(t)}{\partial t} \Big | < \Big |\frac{\partial T_{3}(t)}{\partial t}\Big | < \Big |\frac{\partial T_{2}(t)}{\partial t}\Big |.
\end{equation} 
For a deeper understanding of these observations, let us consider the scenario depicted in the first panel of Fig.~\ref{Fig:H_int_111-000}. By examining the rate of temperature change for each qubit for this case, as shown in Fig.~\ref{Fig:rate_111_000}, we find that the absolute values of the rates $\Big |\frac{\partial T_{j}(t)}{\partial t} \Big |$ for $j=1$, $2$, and $3$, satisfy Eq.~(\ref{eq:rates}) in the transient regime. This fulfillment leads to the realization of an effective QHT (see the first panel of Fig.~\ref{Fig:H_int_111-000}).

Hence, it is evident that for various sets of initial conditions, we can design three-qubit self-contained QHTs with different interaction Hamiltonians. The operation of these QHTs is contingent upon different conditions related to the local temperature or the change in temperature of the qubits over time. This highlights the strength versatility of the three-qubit QHT model, as it can be tailored to construct a perfectly self-contained QHT model for various initial conditions.
We have the flexibility to choose from various configurations for the quantum heat transformer and subsequently select its self-contained interaction Hamiltonian, thus enabling the creation of an intended QHT model.


\begin{figure}[tb]
\includegraphics[width=8.5cm, height=5.8cm]{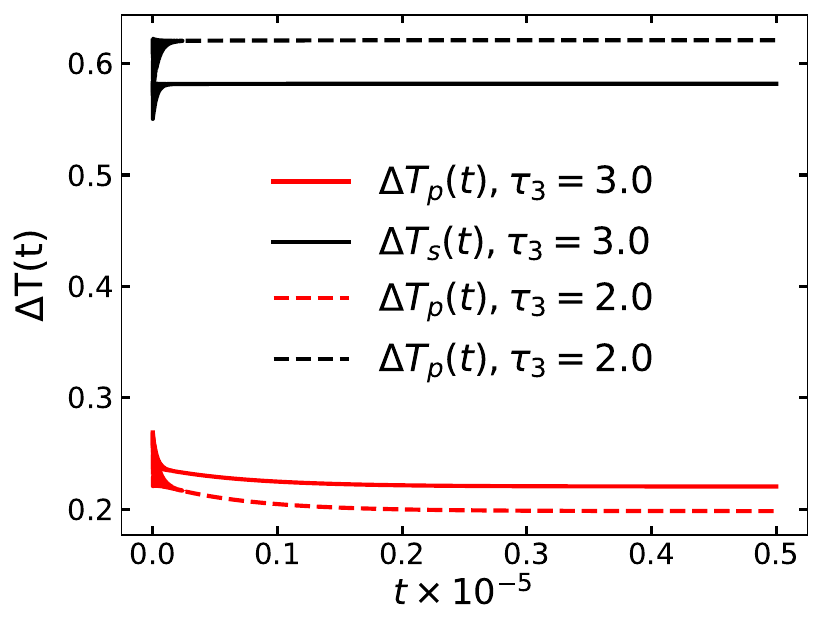}
\caption{Time evolution of temperature gradients at primary and secondary thermal junctions, with the variation in the initial temperature of $q_3$ for the interaction Hamiltonian $\mathcal{H}_{\text{int}}^1$. Here, $\Delta T_p(t)$ is represented by red lines (solid and dashed), while $\Delta T_s(t)$ is denoted by black lines (solid and dashed). As the black curves lie above the red ones, it indicates that the QHTs are operating as step-up transformers. Here, we set $\mathcal{E}_1=-1.0$ and $\mathcal{E}_2=2.0$. The self-contained condition of $\mathcal{H}_{\text{int}}^{1}$ yields $\mathcal{E}_3 = -\mathcal{E}_2 - \mathcal{E}_1$. All other parameters are same as in Fig.~\ref{Fig:H_int_111-000}. All quantities plotted here are dimensionless.}
\label{Fig:step-up_3q}
\end{figure}

\subsubsection{The step-up mode}
The operational characteristics of the QHT in the step-up mode are illustrated in Fig.~\ref{Fig:step-up_3q}, where the interaction between the qubits is governed by the Hamiltonian $\mathcal{H}_{\text{int}}^1$. In this case, the temperature gradients of the secondary junction lie above those of the primary junction, indicating that the QHT operates in the step-up mode within the chosen parameter regime. This behavior is observed consistently across different values of the bath temperature of the third qubit, $\tau_3$.

By comparing Fig.~\ref{Fig:H_int_111-000} and Fig.~\ref{Fig:step-up_3q}, we observe that reversing the sign of the parameter $\mathcal{E}_1$, while keeping all other parameters unchanged, leads to a transition in the operation of the QHT from step-down to step-up mode. A similar transition is observed upon changing the sign of $\mathcal{E}_2$. This indicates that the operational mode of the QHT is highly sensitive to the intrinsic properties of the system, particularly the signs and magnitudes of the energy parameters. To gain deeper insight into this behavior, we now turn to an analysis of how the performance ratio $\mathcal{R}$ varies with these system parameters.


\subsubsection{The performance ratio and its analogy with classical voltage transformer}
\label{Sec:PR}
\begin{figure}[tb]
\includegraphics[scale=0.56]{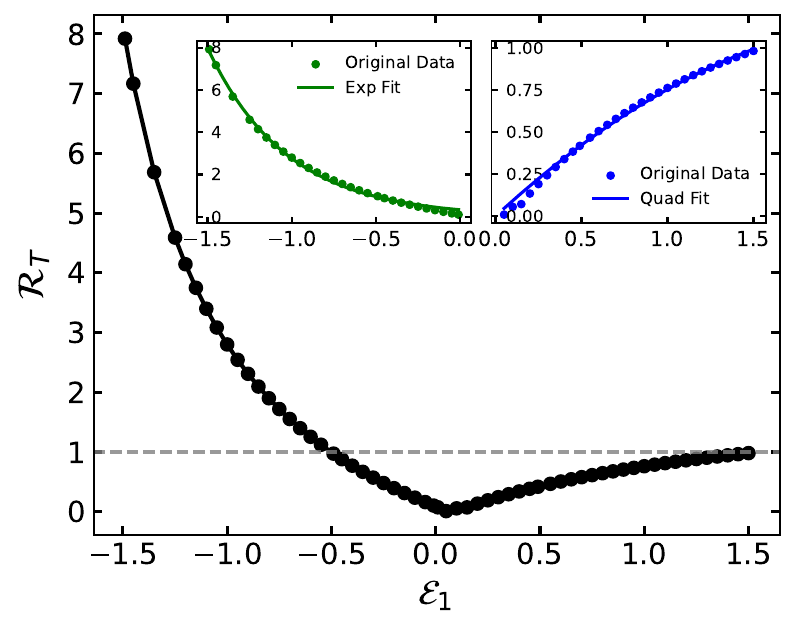}
\caption{ Dependence of the performance ratio $\mathcal{R}_T = \Delta T_s / \Delta T_p$ of the QHT on the parameter $\mathcal{E}_1$ in both the step-up and step-down regimes for the interaction Hamiltonian $\mathcal{H}_{\text{int}}^1$. Here, all parameters are fixed as in Fig.~\ref{Fig:H_int_111-000}. The time is fixed at $t = 0.5 \times 10^5$ and $\tau_3=3.0$. The condition $\mathcal{R}_T < 1$ corresponds to the step-down mode, while $\mathcal{R}_T > 1$ indicates the step-up mode. The dashed black line at $\mathcal{R}_T=1$ serves as the boundary separating the two modes. 
The data in the step-up and step-down regions are well fitted by exponential and polynomial functions, respectively, as detailed in the main text. These fits are also shown separately in the left and right panels of the inset plots. All the quantities plotted here are dimensionless.
}
\label{Fig:E1_rat}
\end{figure}

In Fig.~\ref{Fig:E1_rat}, we study the performance ratio $\mathcal{R}_T$ of a QHT for the interaction Hamiltonian $\mathcal{H}_{\text{int}}^1$ under the self-contained condition $\mathcal{E}_3 = -\mathcal{E}_2 - \mathcal{E}_1$, across the range $-1.5 \leq \mathcal{E}_1 \leq 1.5$. Our results reveal that for $-0.5 \lessapprox \mathcal{E}_1 \lessapprox 1.5  $, the transformer operates in a step-down mode (i.e., $\mathcal{R}_T < 1$), whereas for $\mathcal{E}_1 \lessapprox -0.5 $, it enters a step-up mode (i.e., $\mathcal{R}_T > 1$). This switching between step-up and step-down operation modes can also be achieved by varying $\mathcal{E}_2$, while keeping the other parameters unchanged.

To further understand and optimize the performance within each operational regime, we analyze the functional dependence of $\mathcal{R}_T$ on $\mathcal{E}_1$. 
In the regime $-1.5 \leq \mathcal{E}_1 < 0$, the ratio follows an exponential behavior
\begin{eqnarray*}
\mathcal{R}_T&=& \alpha \exp(\beta \mathcal{E}_1),\ \text{with}\;\; \alpha = 0.330 \pm 0.010 \\
&&\phantom{ami shunech}\text{and}\;\; \beta =-2.123 \pm 0.025.
\end{eqnarray*}
 In the regime $0 < \mathcal{E}_1 \leq 1.5$, $\mathcal{R}_T$ follows a polynomial form
    \begin{eqnarray*}
    \mathcal{R}_T&=& a\mathcal{E}_{1}^2 + b \mathcal{E}_1,\ \text{with}\;\; a =-0.184 \pm 0.014\\
    &&\phantom{ami shunec}\text{and}\;\; b =0.938 \pm 0.017.
    \end{eqnarray*}
See the insets of Fig.~\ref{Fig:E1_rat} for the fitting curves. The error bars represent the standard errors of the fitted parameters, obtained from the fitting procedure under the assumption of independent data points with Gaussian noise. Therefore, depending on the chosen operation mode, the transformer performance can be effectively tuned via the input parameter $\mathcal{E}_1$: exponentially in the $-1.5\le \mathcal{E}_1<0$ regime and polynomially in the $0<\mathcal{E}_1\le 1.5$ regime.

Now, as we know, the voltage ratio in an electrical transformer follows the relation $\mathcal{R}_v=V_{s}/V_{p} = N_{s}/N_{p}$, where $V_s$ and $V_p$ are the voltage differences across the secondary and primary coils, respectively, and $N_s$ and $N_p$ are the corresponding number of coil turns. By tuning the right-hand side ratio of this equation, one can manipulate the voltage ratio on the left-hand side. A value $\mathcal{R}_v<1$ indicates step-down operation, while  $\mathcal{R}_v>1$ corresponds to step-up operation.
In our quantum setting, a similar control mechanism emerges. Here, the tunable parameters are not coil turns but the intrinsic energy parameters of the qubits. Just as classical transformers achieve performance control through mechanical design, quantum heat transformers enable performance tuning, such as switching between step-up and step-down modes, by varying a single system parameter while keeping others fixed.

Note that these relations of $\mathcal{R}_T$ with $\mathcal{E}_1$ are specific to the set of values of the other parameters chosen in Fig.~\ref{Fig:E1_rat}. While the functional forms—exponential in the step-up regime and polynomial in the step-down regime—capture the qualitative behavior within this parameter space, the exact coefficients and the transition point between the two regimes may vary with different choices of system parameters such as qubit energies, bath temperatures, or interaction strengths. 

It is important to note that, unlike an ideal classical voltage transformer that conserves power with perfect efficiency, the quantum heat transformer operates within the framework of open quantum dynamics, where interactions with the environment inevitably lead to decoherence and dissipation. These effects complicate the formulation of a direct analogue to energy or power conservation in the quantum regime.
Despite this, a QHT can still be meaningfully called a transformer as it is capable of functioning in both step-up (see Fig.~\ref{Fig:step-up_3q}) and step-down (see Fig.~\ref{Fig:H_int_111-000}) modes. This transformative behavior clearly distinguishes it from purely dissipative elements like thermal resistors or rheostats, which can only reduce energy flow (e.g., voltage or heat) and cannot perform bidirectional transformation.

In our QHT model, we have considered thermal baths at different temperatures. One may now naturally ask whether the transformation phenomenon can still occur if all the baths are maintained at the same temperature or if only a single thermal bath is used. We investigate this scenario and find that even with same bath temperatures, the system can exhibit both step-up and step-down transformer behavior, although the performance of the transformer is significantly reduced. For this reason, we chose to focus our analysis on configurations with distinct bath temperatures, which allow the machine to operate more effectively in both step-up and step-down regimes. A detailed discussion and analysis of the same-temperature bath scenario are presented in App.~\ref{Sec:same_temp}.

\subsubsection{The behavior of the heat currents}

It is also important to discuss the decoherence-induced energy exchange between the system and the baths. In our framework, energy changes originate from the dissipative part of the GKSL master equation, which captures the irreversible system-environment interaction responsible for decoherence. The corresponding energy flow between the composite three-qubit system and the baths is quantified by the heat current
\begin{equation}
\dot{Q}(t)= \frac{\hbar}{\mathcal{K}}\mathrm{Tr}\!\left[\,\mathscr{H}\, \mathcal{L}(\rho(t))\,\right],
\label{eq:Q}
\end{equation}
where  $\mathscr{H}=\mathscr{H}_{0}+\mathscr{H}_{\text{int}}$, and $\dot{Q}(t)$ is a dimensionless quantity. The sign of $\dot{Q}(t)$ directly indicates the direction of energy flow: $\dot{Q}(t)>0$ corresponds to energy entering the system from the bath, whereas $\dot{Q}(t)<0$ corresponds to energy loss from the system to the bath. Local heat currents are defined as $\dot{Q}_j(t)=\frac{\hbar}{\mathcal{K}}\Tr[\mathscr{H} \mathcal{L}_j(\rho(t))]$.
For the cases shown in Figs.~\ref{Fig:H_int_111-000} and~\ref{Fig:step-up_3q} with $\tau_3 = 3.0$, the behavior of both the total and local heat currents is presented in Fig.~\ref{fig:Q-diagram} in App.~\ref{app:heat_current}, where we also discuss the relation between the local heat currents and the corresponding local temperatures. In many quantum thermal devices, such heat-current quantities directly characterize performance. For example, the cooling power of a quantum refrigerator, the amplification factor of a thermal transistor, or the rectification ratio of a quantum heat diode~\cite{Levy1,Joulain,Guo_2018,Liu_2024}.

   \begin{table}[htb]
\centering
\begin{tabular}{c|c|c|c}
$\tau_3$ & $ \dot{Q}_{ p}\times 10^6$ & $ \dot{Q}_{ s}\times 10^6$ & Analysis \\
\hline
\multicolumn{4}{c}{\textbf{In step-down mode Conditions}} \\
\hline
3.0 & -25.681 & 10.718 & Primary loss rate $>$ Secondary gain rate \\
2.5 & -19.805 & 8.423  & Primary loss rate $>$ Secondary gain rate \\
2.0 & -11.722 & 5.264 &Primary loss rate $>$ Secondary gain rate \\
\hline
\multicolumn{4}{c}{\textbf{In step-up mode Conditions}} \\
\hline
3.0 & -1.139 & 5.243 & Primary loss rate $<$ Secondary gain rate \\
2.5 & -0.283  & 5.239  & Primary loss rate $<$  Secondary gain rate \\
2.0 & 0.925 & 5.240  & Primary gain rate $<$ Secondary gain rate \\
\end{tabular}
\caption{Heat currents flowing into or out of the primary ($\dot{Q}_p$) and secondary ($\dot{Q}_s$) junction at time $t=0.5 \times 10^{5}$ for different values of $\tau_3$. All other parameters are same as in Figs.~\ref{Fig:H_int_111-000} and~\ref{Fig:step-up_3q} for the step-down and step-up mode conditions, respectively. The “Analysis’’ column summarizes the relative magnitudes of heat gain and loss rate.
All the quantities whose numerical values are presented here are dimensionless.
}
\label{Ref3:Tab:Q-dot}
\end{table}

\begin{figure}[htb]
	\includegraphics[scale=0.50]{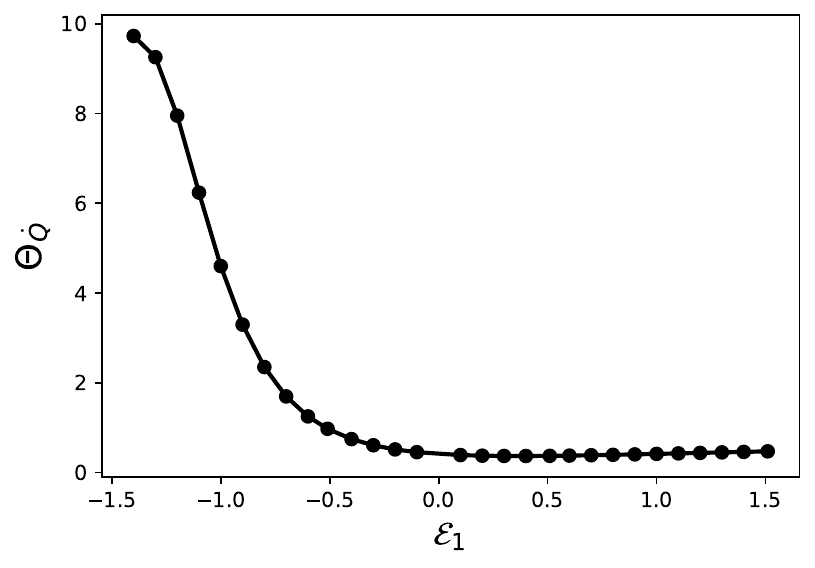}
        \caption{The behavior of the heat current ratio $\Theta_{\dot{Q}}=|\dot{Q}_{s}|/|\dot{Q}_{p}|$ of a QHT as a function of the energy gap $\mathcal{E}_{1}$  for the interaction Hamiltonian $\mathcal{H}_{\text{int}}^1$. All the parameters are same as in the Fig.~\ref{Fig:H_int_111-000}. Here, $\tau_3=3.0$ and time $t=0.5\times 10^5$. All the quantities plotted here are dimensionless.}
        \label{Fig:Ref3_R_Q_dot}
	\end{figure}

To understand the dependence of the performance of a quantum heat transformer on the heat flow between the system and the thermal baths, we define the instantaneous heat current flowing into or out of the primary and secondary junctions as $\dot{Q}_{p}(t)$  = $\dot{Q}_1(t) + \dot{Q}_2(t)$ and $\dot{Q}_{s}(t)$ = $\dot{Q}_2(t) + \dot{Q}_3(t)$, respectively, where a positive value indicates heat absorption rate at time $t$, and a negative value indicates the rate of the heat release. 
In Table~\ref{Ref3:Tab:Q-dot}, we present the heat currents flowing into or out of the primary and secondary junctions for the parameter choices used in Fig.~\ref{Fig:H_int_111-000} (step-down mode conditions) and Fig.~\ref{Fig:step-up_3q} (step-up mode conditions), evaluated at $t=0.5\times 10^5$. From Table~\ref{Ref3:Tab:Q-dot}, we observe that in the step-down mode the magnitude of the heat change rate at the primary junction exceeds that at the secondary junction, whereas in the step-up mode the heat change rate at the primary junction is smaller than that at the secondary junction. Consequently, defining the heat-current ratio of the junctions as $\Theta_{\dot{Q}}=|\dot{Q}_{s}|/|\dot{Q}_{p}|$, we find that $\Theta_{\dot{Q}} < 1$ corresponds to the step-down mode, while $ \Theta_{\dot{Q}} >1 $ indicates the step-up mode for the scenarios presented in Table~\ref{Ref3:Tab:Q-dot}. The behavior of $\Theta_{\dot{Q}}$ with the system parameter $\mathcal{E}_1$ for $\tau_3=3.0$ and $t=0.5 \times 10^5$ is shown in Fig.~\ref{Fig:Ref3_R_Q_dot}. The absolute values are used because the heat change rate may represent either heat gain or heat loss, and a junction may switch between absorbing and releasing heat across different parameter regimes within the same operational mode. For example, for $\tau_3=2.0$ under the step-up conditions, the primary junction absorbs heat, whereas for other $\tau_3$ values in the step-up regime it releases heat.

If we now consider the total heat absorbed or released by the junctions up to a given time, instead of the instantaneous heat currents, we observe behavior consistent with that shown in Table \ref{Ref3:Tab:Q-dot}. The total heat exchanged by the $j$th qubit over an evolution time $t_f$ is defined as $Q_j^{ \text{tot}} = \int_{0}^{t_f} \dot{Q}_j(t)\, dt$. A detailed discussion on this matter is provided in App.~\ref{app:tot_q}.


 In App.~\ref{app:laws} we verify the consistency of the QHT dynamics with the first and second laws of thermodynamics: the first law follows from energy conservation, linking the change in internal energy to the net heat currents, while the second law is ensured via Spohn’s theorem, which guarantees non-negative entropy production under completely positive Markovian dynamics.

\begin{figure*}
\includegraphics[width=5.9cm, height=4.5cm]{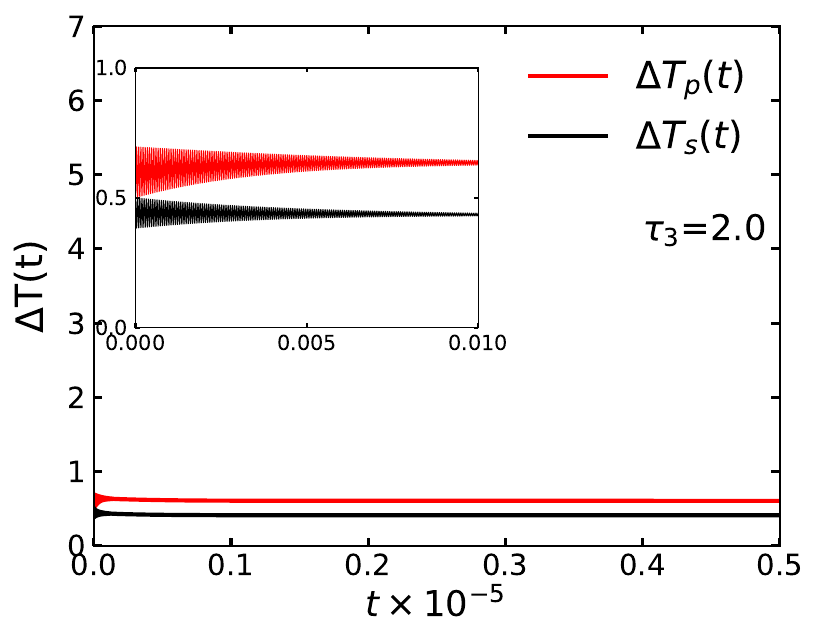}
\includegraphics[width=5.9cm, height=4.5cm]{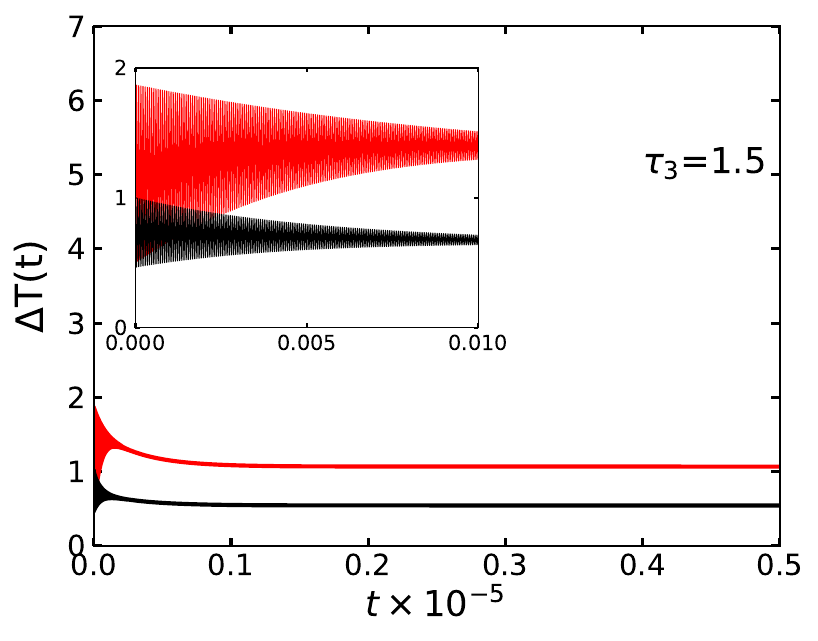}
\includegraphics[width=5.9cm, height=4.5cm]{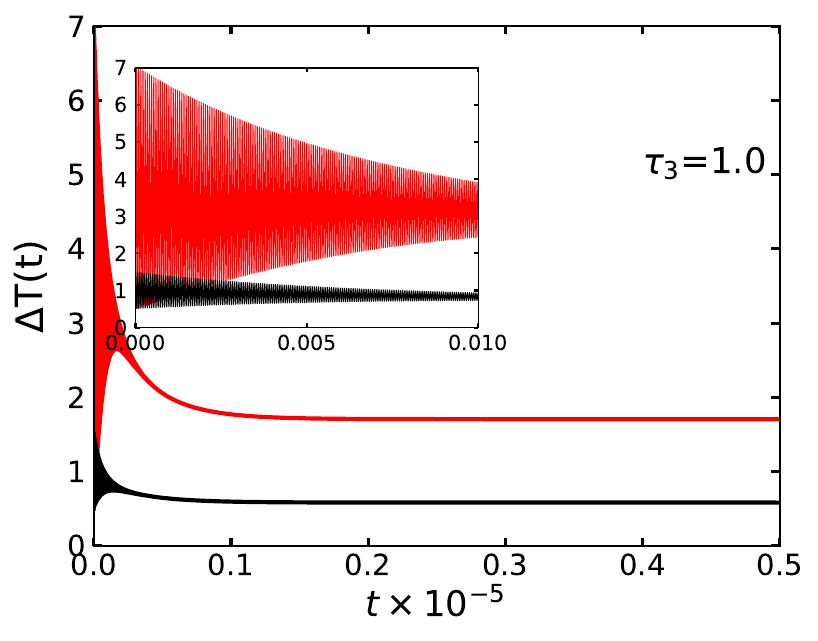}
\caption{Time dynamics of temperature gradients at primary and secondary thermal junctions, with the variation in the initial temperature of $q_3$ for the interaction Hamiltonian $\mathcal{H}_{\text{int}}^2$. Here, the quantities, $\Delta T_p(t)$ (red line) and $\Delta T_s(t)$ (black line), are plotted with respect to $t$ for different values of $\tau_3$. As the red curves lie above the black ones, it indicates that the QHTs are operating as step-down transformers.
For this case, the self-contained condition is given by $\mathcal{E}_3=\mathcal{E}_2-\mathcal{E}_1$. The initial temperatures of $q_1$ and $q_2$  
are taken as $\tau_1=3.0$, $\tau_2 =2.5$. All other parameters are same as in Fig.~\ref{Fig:H_int_111-000}. All the quantities plotted here are dimensionless.}
\label{Fig:101_010}
\end{figure*}

\subsection{QHT model: The interaction Hamiltonian $\mathcal{H}_{\text{int}}^{2}$}
Now, for completeness, we proceed to discuss another example of a QHT model involving the interaction Hamiltonian $\mathcal{H}_{\text{int}}^{2}$, which has already been briefly introduced earlier.
For this case also, the transition frequencies are the same as in the case of $\mathcal{H}^1_{\text{int}}$.
In this QHT model, we take the initial temperatures of the three qubits in the order $\tau_{1} > \tau_{2} > \tau_{3}$. As previously mentioned, we configure the parameters of the systems and the interaction Hamiltonians such that, $\mathcal{P}^{\text{in}}_{101}>\mathcal{P}^{\text{in}}_{010}$, with $\mathcal{P}^{\text{in}}_{101}=(1-p_{1}^{\text{in}})p_{2}^{\text{in}}(1-p_{3}^{\text{in}})$ and $\mathcal{P}^{\text{in}}_{010}=p_{1}^{\text{in}} (1-p_{2}^{\text{in}}) p_{3}^{\text{in}}$. 
In Fig.~\ref{Fig:101_010}, we present the operational characteristics of the QHT for this configuration. 
The temperature gradients of the primary and secondary junctions, $\Delta T_p(t)$ and $\Delta T_s(t)$, exhibit a different behavior compared to the case described in Fig.~\ref{Fig:H_int_111-000}. Initially, both $\Delta T_p(t)$ and $\Delta T_s(t)$ oscillate,and then gradually decrease to reach a steady-state value over time. Furthermore, as the initial temperature of $q_3$ decreases, the steady-state difference between $\Delta T_p(t)$ and $\Delta T_s(t)$ increases. 
This behavior contrasts with the scenario depicted in Fig.~\ref{Fig:H_int_111-000}.


So far, our investigation has focused on cases where the QHT models function as step-down transformers in both steady-state and transient regimes. These models are undeniably crucial for designing effective step-down quantum transformers. However, in certain scenarios, the initial conditions may impose constraints such that they yield a steady-state step-up transformer, whereas the requirement is for a step-down operation. In such cases, a necessarily transient step-down transformer model can prove to be beneficial. The term ``necessarily transient step-down transformer" implies that the desired step-down mode can be achieved within the transient domain in an steady-state step-up QHT model. These models hold significance, particularly for specific initial conditions. In the next section, we will present an example of a necessarily transient step-down quantum heat transformer and delve into its operational characteristics. 
\section{Necessarily Transient quantum heat transformers}
\label{Sec:3}
\begin{figure}
\includegraphics[width=8.5cm, height=5.8cm]{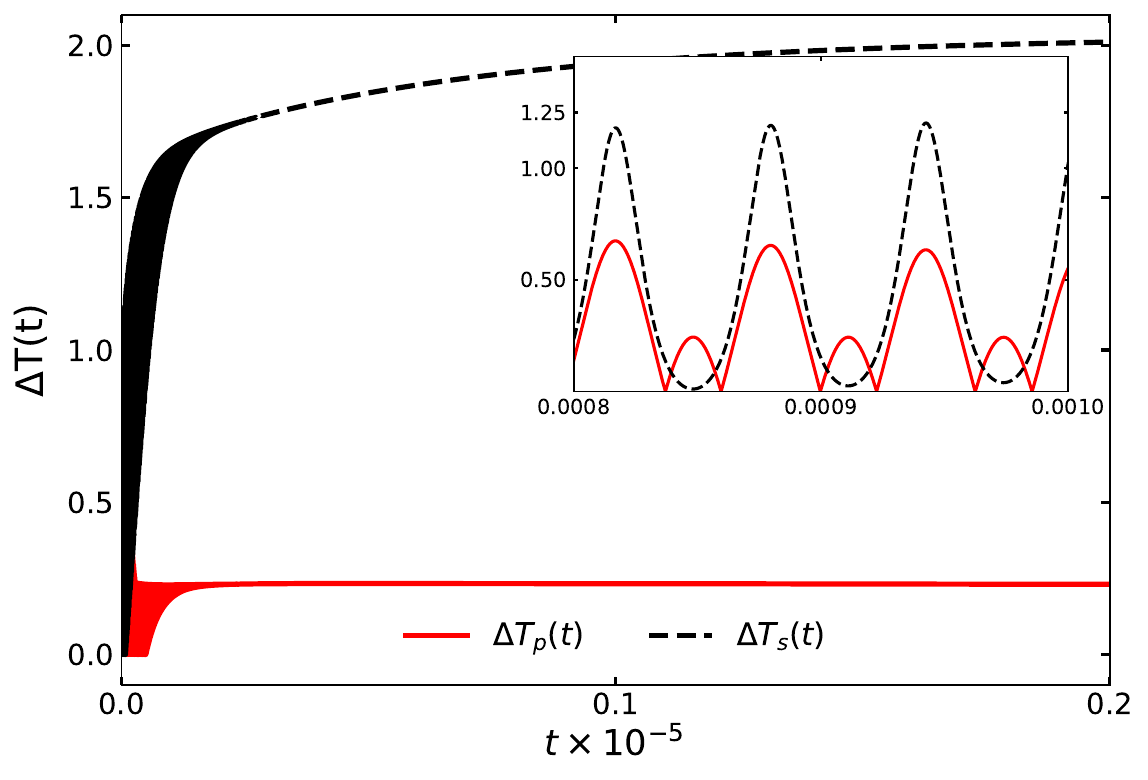}
\caption{Necessarily transient step-down quantum heat transformer. Here, we plot $\Delta T_p(t)$ with the red line and $\Delta T_s(t)$ with the black line with respect to time for the interaction Hamiltonian 
$H_{\text{int}}^{2}$. The necessarily transient step-down operation of the transformer is shown in the inset. The regions where the red curve lies above the black curve indicate the regime of necessarily transient step-down operation. Here we take, $\mathcal{E}_1 = 2.0$, $\mathcal{E}_2 = 1.4$, $\tau_1=1.0$, $\tau_2=2.0$, and $\tau_3 = 3.0$. All other considerations are same as in Fig.~\ref{Fig:H_int_111-000}. 
All the quantities plotted here are dimensionless.}
\label{Fig:101_NT}
\end{figure}
Let us again consider the QHT protocol with the $\mathcal{H}_{\text{int}}^{2}$ interaction Hamiltonian. As we intend to model a step-up transformer in the steady-state regime, in this model, we need to take the initial temperatures of the three qubits in reverse order compared to the previous step-down configuration. Therefore, it follows: $\tau_{1} < \tau_{2} < \tau_{3}$. In this case also, initially we fix the parameters such that $\mathcal{P}^{\text{in}}_{101}> \mathcal{P}_{010}^{\text{in}}$. 
We set the initial temperatures of qubits with equal differences, i.e., at time $t=0$ we set, $\Delta T_{p}^{i} = \Delta T_{s}^{i} = 1.0 $.
The time dynamics of $\Delta T_p(t)$ and $\Delta T_s(t)$ is shown in Fig.~\ref{Fig:101_NT} for this situation. We observe that, starting from an equal value, the temperature difference in the primary junction $\Delta T_{p}(t)$ decreases with time. On the other hand, the temperature difference of the secondary junction,  $\Delta T_{s}(t)$ increases with time, and the protocol indicates a step-up transformer overall. However, an intriguing discovery emerges upon closer inspection of very small time values. Within this time domain, it becomes apparent that the temperature at the secondary junction falls below that of the primary junction. See the inset of Fig.~\ref{Fig:101_NT}.
This observation suggests a step-down configuration, albeit exclusively within a transient domain. We term these QHT models as ``the necessarily transient step-down QHT". This discovery is unique within the QHT protocol, as it reveals that a step-up transformer can exhibit a step-down behavior in the transient domain under specific configurations.

The motivation for studying the necessarily transient operation of QHTs stems from practical considerations in quantum technologies. In realistic scenarios, waiting for a device to reach the steady state can pose challenges, especially since qubits are prone to decoherence and noise from the surrounding environment. These effects can significantly disrupt the intended operation if the steady-state timescale is too long. To address this, recent research has emphasized the advantages of transient operation, where the device performs its function within a shorter time window before reaching equilibrium (see~\cite{Brask2015,Mitchison_2015,Das_2019,Ghosh2021,saha2023,Maity2024} for examples in other quantum thermal devices). Motivated by this, we explore whether the quantum heat transformer can operate effectively in the transient regime. More importantly, we identify parameter regimes where the QHT exhibits necessarily transient behavior—i.e., step-up or step-down thermal transformation occurs only during the transient evolution and disappears in the steady state. In such cases, restricting the analysis to steady-state behavior would entirely miss the functional window of the device, emphasizing the critical role of necessarily transient dynamics in quantum thermal technologies.

\begin{figure}
		\centering
	\includegraphics[scale=0.2]{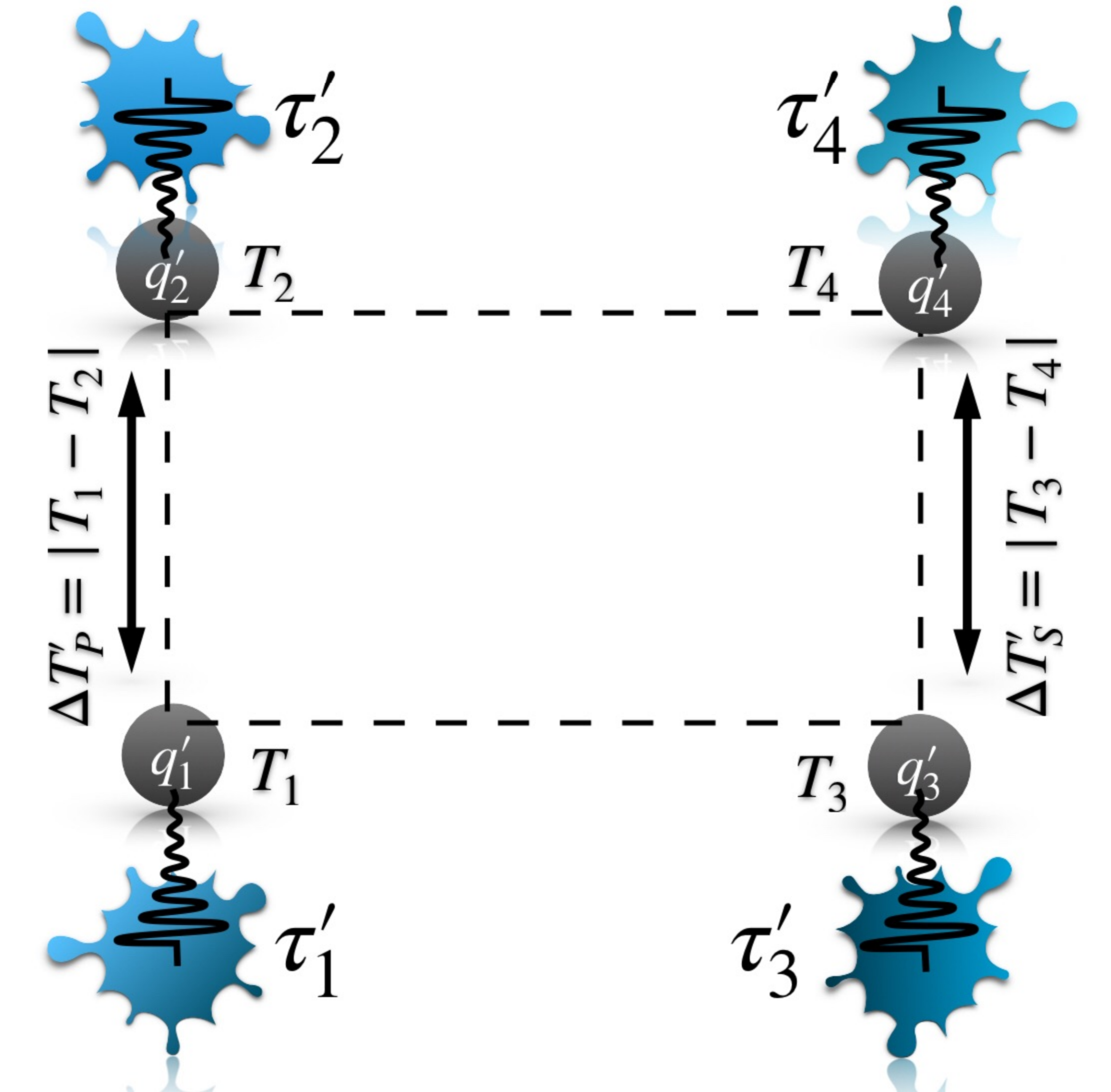}
		\caption{Schematic diagram of a four-qubit QHT. Here, $q_1^{\prime}$, $q_2^{\prime}$, $q_3^{\prime}$, and $q_4^{\prime}$ represent the four qubits of the quantum transformer, each immersed in a separate thermal bath at dimensionless temperatures $\tau_1^{\prime}$, $\tau_2^{\prime}$, $\tau_3^{\prime}$ and $\tau_4^{\prime}$, respectively. The pair of qubits $q_1^{\prime}$ and $q_2^{\prime}$ forms the primary thermal junction, while the pair $q_3^{\prime}$ and $q_4^{\prime}$ constitutes the secondary thermal junction. At any given time $t$, the local temperatures of the qubits are represented as $T_1(t)$, $T_2(t)$, $T_3(t)$ and $T_4(t)$. The temperature gradient across the primary thermal junction is defined as $\Delta T_p^{\prime}(t) = |T_1(t) - T_2(t)|$, while across the secondary junction it is denoted as $\Delta T_s^{\prime}(t) = |T_3(t) - T_4(t)|$.  }
		\label{fig:4qubit_QHT}
	\end{figure}
     \begin{figure*}[!htb]
\includegraphics[width=5.9cm, height=4.5cm]{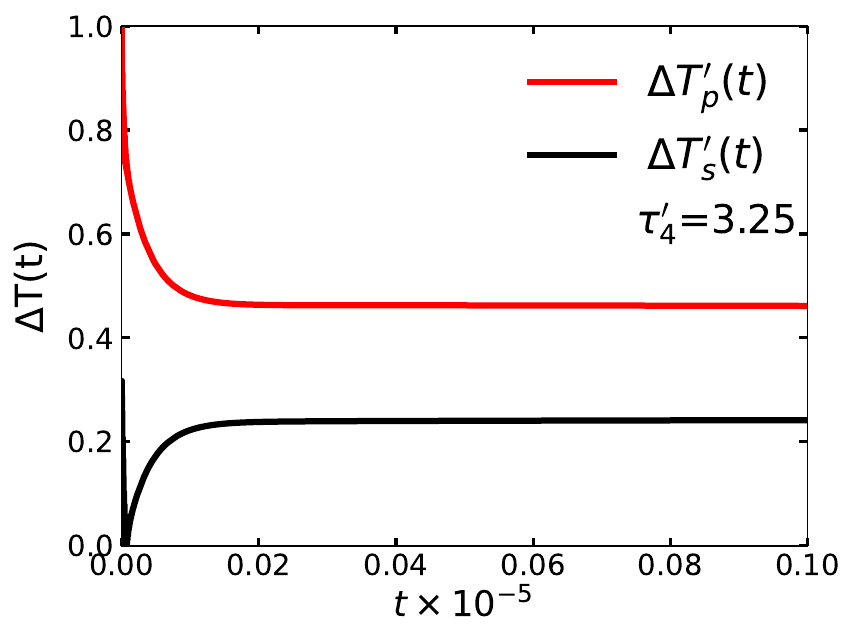}
\includegraphics[width=5.9cm, height=4.5cm]{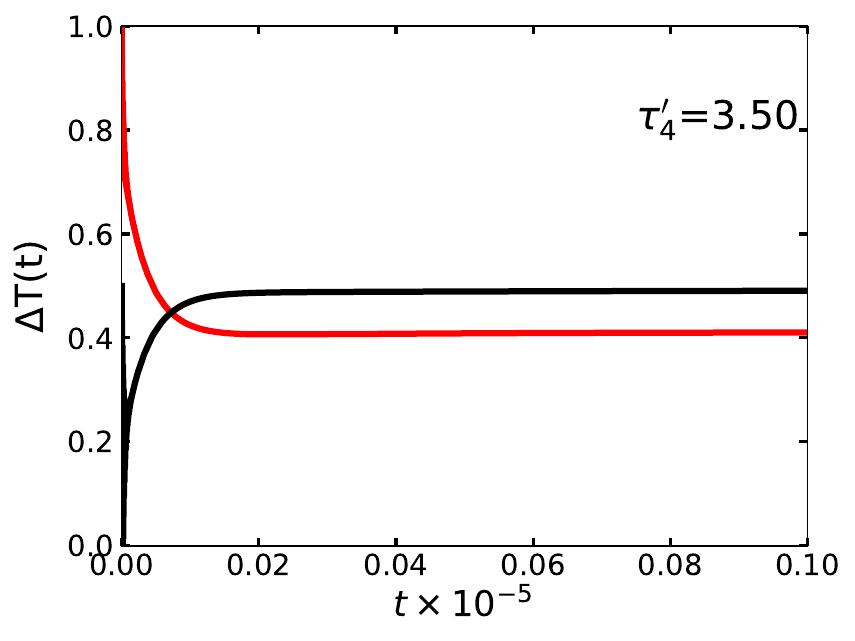}
\includegraphics[width=5.9cm, height=4.5cm]{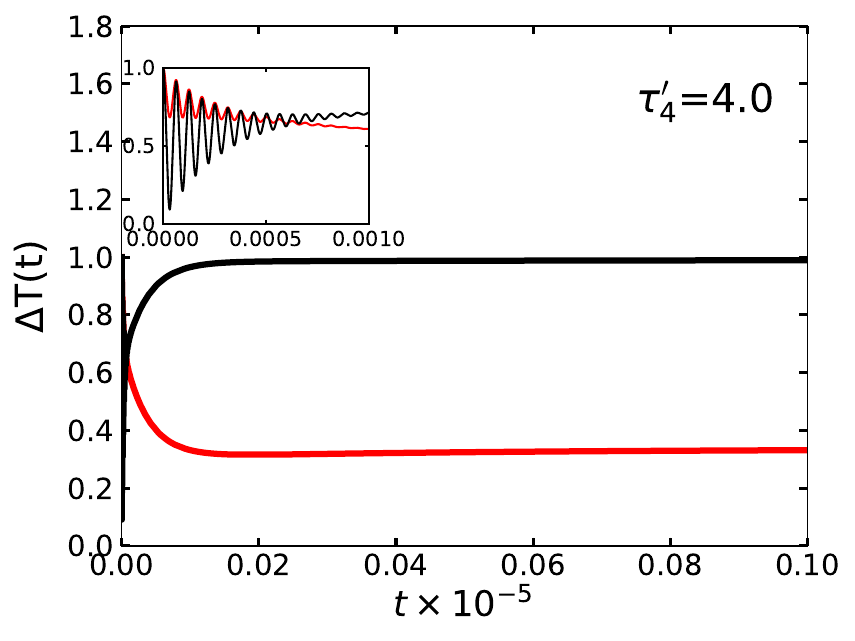}
\caption{Four-qubit QHT: time evolution of temperature gradients at primary and secondary thermal junctions, with the variation in the initial temperature of $q_4$ for the interaction Hamiltonian $\mathcal{H}_{\text{int}}^{\prime}$. Here, the temperature gradient of the primary thermal junction, $\Delta T_p^{\prime}(t)$, and that of the secondary thermal junction, $\Delta T_s^{\prime}(t)$, are plotted with time $t$, for different values of $\tau_4^{\prime}$. The left most panel shows a steady-state step-down QHT, while the two right panels depict steady-state step-up QHTs that exhibit necessarily transient step-down behavior during their initial evolution. The transition region between these two modes for $\tau_4^{\prime}=4.0$ is shown in the inset of the right most panel. The regions where the red curves lie above the black curves indicate the regimes of necessarily transient step-down operation.
The parameters are taken as $\mathcal{E}_1 = 1.0$, $\mathcal{E}_2 = 2.0$, $\mathcal{E}_3 = 3.0$, $\tau_1^{\prime}=1.0$, $\tau_2^{\prime} =2.0$, $\tau_3^{\prime}=3.0$, $g^{\prime}=0.5$, $\alpha_4=10^{-2}$, and $\Omega_4=10^3$. All the other parameters are same as in Fig.~\ref{Fig:H_int_111-000}. 
All quantities plotted here are dimensionless.
}
\label{Fig:4qubit_1010_0101}
\end{figure*}

\section{Self-contained Four-qubit QHT}
\label{Sec:4}
We now move to the four-qubit QHT model consisting of four two level systems, $q_1^{\prime}$, $q_2^{\prime}$, $q_3^{\prime}$, and $q_4^{\prime}$, locally connected to four thermal baths at temperatures $\tau_1^{\prime}$, $\tau_2^{\prime}$, $\tau_3^{\prime}$, and $\tau_4^{\prime}$, respectively. For this model, the primary junction corresponds to qubits $q_1^{\prime}$ and $q_2^{\prime}$, and the secondary junction corresponds to qubits $q_3^{\prime}$ and $q_4^{\prime}$. Therefore, there is no shared qubit between the primary and secondary junctions, unlike in the three-qubit QHT model. Here, the temperature gradient at primary junction is given by $\Delta T_p^{\prime}(t)=|T_1(t)-T_2(t)|$, and the same for the secondary junction is defined as $\Delta T_s^{\prime}(t)=|T_3(t)-T_4(t)|$, with $T_j(t)$, for $j=1$, $2$, $3$, and $4$, being the local temperatures of the four qubits, respectively. A schematic diagram of a four-qubit QHT is depicted in Fig.~\ref{fig:4qubit_QHT}.

The free Hamiltonian of the composite four-qubit system is represented by $\mathcal{H}_{0}^{\prime} = \frac{\mathcal{K}}{2} \sum_{j=1}^{4} \mathcal{E}_{j}\sigma_{j}^{z}$. 
In a similar manner to the three-qubit system, , various interaction Hamiltonians can also be formulated here, each accompanied by its respective self-contained conditions.
Instead of delving into all potential interaction Hamiltonians, we focus on a single example to examine various properties of the four-qubit QHT model. The chosen interaction Hamiltonian and the associated self-contained condition, for this four-qubit model, is given by
\begin{equation}
    \mathcal{H}_{\text{int}}^{\prime} = \mathcal{K}g^{\prime}(|1010\rangle \langle 0101 |) + h.c ~; ~ ~ \mathcal{E}_4 = \mathcal{E}_3 - \mathcal{E}_2 + \mathcal{E}_1.
\end{equation}
Here, $g^{\prime}$ is the dimensionless coupling strength. The initial state of  the four-qubit system is taken as $\rho^{\prime}(0) = \bigotimes_{j=1}^{4} \rho_{j}(0) $, for $j=1$ to $4$. We set the parameters $\mathcal{E}_j$s and $\tau_j$s for $j=1$ to $4$, such that
\begin{align}
\mathcal{P}^{\text{in}}_{1010} &> \mathcal{P}^{\text{in}}_{0101} \nonumber \\
\implies  (1-p_{1}^{\text{in}})p_{2}^{\text{in}}(1-p_{3}^{\text{in}})p_{4}^{\text{in}} & > p_{1}^{\text{in}}(1-p_{2}^{\text{in}})p_{3}^{\text{in}}(1-p_{4}^{\text{in}}).
    \label{kichhu_ekta}
 \end{align}
This implies that the initial probability of finding the system in the state $\ket{1010}$ is higher than that of $\ket{0101}$. Similar to the three-qubit case, we consider the thermal baths as bosonic baths described by the Hamiltonian given in Eq.~(\ref{eq:Mar_bath}), with system-bath interactions represented by Eq.~(\ref{eq:bath_int}) for $j=1$ to $4$.
We now allow the system to evolve under the Markovian dynamics, governed by Eq.~(\ref{Equ:Master_Equ}). For this scenario, the transition frequencies are $\omega \in \{\mathcal{E}_{1},\,\mathcal{E}_{1}\pm g,\; \mathcal{E}_{2},\,\mathcal{E}_{2}\pm g,\; \mathcal{E}_{3},\,\mathcal{E}_{3}\pm g, \mathcal{E}_4, \mathcal{E}_4 \pm g\}$. The time dynamics of $\Delta T_p^{\prime}(t)$ and $\Delta T_s^{\prime}(t)$ is demonstrated in Fig.~\ref{Fig:4qubit_1010_0101}. We examine these quantities while varying the initial temperature of $q_4$. For the case depicted in the first panel of Fig.~\ref{Fig:4qubit_1010_0101}, i.e., for $\tau_4^{\prime}=3.25$, we observe that the heat difference in the primary junction, $\Delta T_{p}^{\prime}(t)$, decreases with time. Simultaneously, the heat difference in the secondary junction, $\Delta T_{s}^{\prime}(t)$, increases with time, ultimately manifesting a step-down behavior. However, in the other two panels, the QHTs exhibit both the nature of a steady-state step-up quantum transformer and the nature of a necessarily transient step-down quantum transformer. For the second case, for $\tau_4^{\prime}=3.5$, we observe that up to a certain threshold value, the transformer exhibits a step-down behavior, but beyond this threshold, it switches to a step-up mode. With a further increase in $\tau_4^{\prime}$ from $3.5$ to $4.0$, the temperature in the secondary junction increases at a faster rate, while the temperature in the primary junction decreases significantly at a faster rate as well. As a result, the step-down mode persists for a very short period of time, and the protocol predominantly functions as a step-up transformer. Therefore, in this four-qubit QHT model, the protocol can be designed to exhibit both step-up and step-down behavior with the same setup, by only varying the initial temperature of $q_4^{\prime}$, and for fixed initial conditions, we can achieve both step-down and step-up operations in different time regions.
This dual nature can also be attained by appropriately adjusting the initial temperature of any one of the qubits. This capability not only expands the versatility of the QHT model but also reduces the design complexity and cost associated with implementing different setups.
Note that, 
in the three-qubit protocols, we can also observe the dual nature discussed in the necessarily transient study. However, the operating region of this necessarily transient phenomenon is very limited, and extending this region is not easily achievable like four-qubit protocol.




Until now, we have discussed the smallest realizations of a quantum heat transformer, namely the three-qubit configuration, and extended our investigation to four-qubit QHTs, where defining two thermal junctions—primary and secondary—is relatively straightforward. In these minimal models, each junction is composed of exactly two qubits, with one qubit assigned to each of the two units that form a junction. However, to extend the QHT concept to systems with more than four qubits, a more general protocol is necessary. In this extended framework, each thermal junction still consists of two distinct units, but each unit can now be made up of one or more qubits. For instance, in a five-qubit system, one unit of a junction consists of two qubits grouped together, while the other unit is to be a single qubit. The remaining qubits form the second junction. The temperature of each unit is then defined as the average of the local temperatures of the qubits it contains. The temperature gradient across a junction is given by the difference between the average temperatures of its two units. This protocol preserves the structural analogy to classical transformers while allowing scalability to larger systems. This approach enables us to generalize the QHT framework to any finite number of qubits, although the associated computational complexity grows rapidly with system size.

Note that, we have restricted our analysis to the Markovian regime, which allows us to employ the GKSL master equation to describe the evolution of the system. In contrast, studying non-Markovian dynamics would require a completely different methodological framework. Such analyses are considerably more complex, especially under strong coupling conditions, where the definition of quantities such as temperature becomes highly nontrivial. For most non-Markovian processes, the reduced state of the system is not guaranteed to remain diagonal in time due to information backflow, making the notion of a local qubit temperature at time $t$ significantly more difficult to define. Consequently, any investigation of QHT performance in non-Markovian regimes would first require establishing consistent and physically meaningful definitions of these quantities before analyzing the operational properties of QHTs.

\section{Conclusion}
\label{Sec:con}
In summary, we have explored the concept and design of quantum heat transformers as a new class of quantum thermal devices to actively manage heat flow in quantum circuits. Inspired by classical heat transformers and electrical voltage transformers, our study highlights how QHTs could redistribute heat to reduce decoherence. 
Operating between two thermal junctions, the QHT modulates temperature gradients in a manner analogous to classical absorption heat transformers, functioning without external work input. 
To explore the operational characteristics of the quantum heat transformer, we focus on the step-down mode as the main aspect of our study. Initially, we investigate the three-qubit QHT model, serving as the smallest quantum heat transformer configuration. Our findings demonstrate that under various interaction Hamiltonians between the three qubits, the model can function as a self-contained quantum heat transformer.
The operation of the QHT protocol relies on the transitions of the qubits between their ground and excited states during the evolution of the system, which impacts the increase or decrease of the local temperatures of the qubits over time. However, in some cases, the behavior of the QHTs cannot be solely explained by changes in local temperatures but also relies on the rate of temperature change of the qubits throughout the evolution of the system. It is important to note that the QHT protocol studied here is analyzed in a generalized manner, allowing for easy switching to the step-up mode as a primary focus. 
We define a key performance indicator of a QHT, the performance ratio ($\mathcal{R}_T$), which measures the capability of both step-up and step-down QHTs. Moreover, we demonstrate that $\mathcal{R}_T$ intrinsically depends on the energy parameters of the system, highlighting a formal analogy with classical voltage transformers, where performance is similarly governed by intrinsic factors such as coil turn ratios.
However, unlike ideal classical transformers, QHTs face inherent limitations due to unavoidable system–environment interactions, which introduce dissipation and make it challenging to formulate a direct
analogue of energy or power conservation in the quantum regime. Despite these challenges, the ability of QHTs to operate in both step-up and step-down modes justifies their designation as ``transformers". This bidirectional functionality clearly distinguishes them from purely dissipative elements such as thermal resistors or rheostats, which can only reduce energy flow and lack the capacity for energy transformation.
Furthermore, we extend our study to four-qubit scenarios, demonstrating the feasibility of constructing self-contained QHTs.
An important discovery of our study is the emergence of the necessarily transient step-down QHT model, observed in both the three and four-qubit protocols. In this  necessarily transient QHT model, a dual-mode nature is evident, wherein the desired step-down mode can be achieved within the transient domain in an originally step-up QHT model. In the three-qubit protocol, a necessarily transient step-down quantum transformer can be obtained for certain qubit interactions and specific initial conditions within a small transient time regime. Conversely, in the four-qubit protocol, the emergence of necessarily transient step-down quantum transformers is more common. In four-qubit scenarios, we have the advantage of easily tuning the transient regime, enabling the transformer to operate in the intended step-down mode simply by controlling the initial temperatures of the qubits within the same setup.
This presents a significant advantage in cost reduction, eliminating the need for distinct setups to achieve step-up and step-down modes.
Hence, this quantum heat transformer model not only serves as an analog to the classical transformers model but also exhibits advanced characteristics, allowing it to operate as both a step-up and step-down transformer within the same setup, a capability absent in classical voltage transformers. 

\begin{figure*}[hbt]
\centering
\includegraphics[width=8.5cm, height=5.8cm]{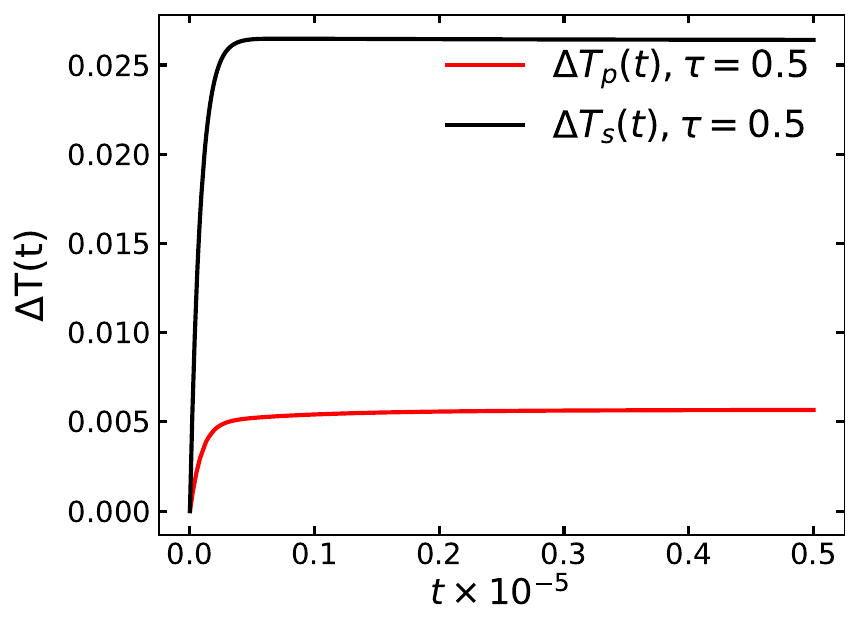}
\includegraphics[width=8.5cm, height=5.8cm]{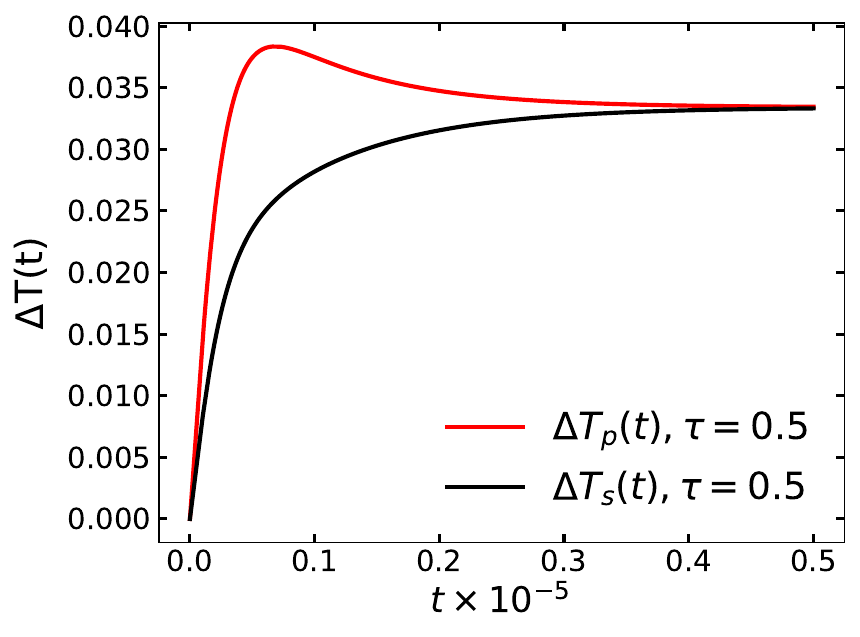}
\caption{Time evolution of temperature gradients at primary and secondary thermal junctions, with the same initial temperature of three baths, $\tau_1=\tau_2=\tau_3=0.5$, for the interaction Hamiltonian $\mathcal{H}_{\text{int}}^1$. In each panel, the temperature gradient of the primary thermal junction, $\Delta T_p(t)$, is represented by red lines, while that of the secondary thermal junction, $\Delta T_s(t)$, is denoted by black lines. 
Here, we set $\mathcal{E}_1=1.0$ and $\mathcal{E}_2=2.0$ for the step-up mode (left Panel) and $\mathcal{E}_1=-1.0$ and $\mathcal{E}_2=2.0$ for the step-down mode (right Panel).
The self-contained condition of $\mathcal{H}_{\text{int}}^{1}$ yields $\mathcal{E}_3 = -\mathcal{E}_2 - \mathcal{E}_1$. All other parameters are the same as in Fig.~\ref{Fig:H_int_111-000}. 
All quantities plotted here are dimensionless.
}
\label{Fig:same_temp}
\end{figure*}


For these reasons, QHTs present a promising tool for thermal management in quantum processors, nanoscale devices, and next-generation energy applications, where precise thermal control is essential for maintaining stability, minimizing thermal noise, and enhancing overall efficiency. In particular, QHTs could help redistribute heat across processors, ensuring uniform temperature distribution and preventing thermal stress on critical components. Their self-contained nature, requiring no external power source, makes them especially suitable for compact nanodevices where energy resources are limited. 
There are several notable experimental realizations of self-contained autonomous thermodynamic devices in the quantum and nanoscale regimes~\cite{Huang_PRL_2024,Maslennikov_Nature_2019,Aamir_Nature_2025}.   They provide experimental validation of such self-contained models using nuclear spin systems, trapped ions, and superconducting circuits. Similar physical platforms could be employed to realize the QHT model proposed in this work.
In this way, QHTs provide a framework for managing heat at the quantum level, enabling more efficient and reliable operation of emerging quantum technologies, with scope for further improvement through diverse implementations. 

\acknowledgements 
 We acknowledge computations performed using Armadillo~\cite{Sanderson,Sanderson1}. We also acknowledge the use of \href{https://github.com/titaschanda/QIClib}{QIClib} -- a modern C++ library for general purpose quantum information processing and quantum computing (\url{https://titaschanda.github.io/QIClib}) and cluster computing facility at Harish-Chandra Research Institute. 
   A. Ghoshal acknowledges the support from the Alexander von Humboldt Foundation.
   We acknowledge partial support from the Department of Science and Technology, Government of India through the QuEST grant (grant number DST/ICPS/QUST/Theme-3/2019/120).

   \appendix
 \section{The details of the bosonic baths and the system-bath interactions}
\label{appen:1}

In the three-qubit QHT model, let us consider a scenario where the three qubits are interacting through $\mathcal{H}_{\text{int}}^1$, and the three thermal baths are bosonic baths, each consisting of an infinite number of harmonic oscillators. The Hamiltonian of each bath can be expressed as
\begin{equation}
    H_{B_j}=
    \int_{0}^{\Omega_j}\hbar\tilde{\omega}b_{\omega^{\prime}}^{j\dagger}b_{\omega^{\prime}}^jd\omega^{\prime},
    \label{eq:Mar_bath}
\end{equation}
for $j=1$ to $3$. Here the operators $b^j_{\omega^{\prime}}$ ($b^{j\dagger}_{\omega^{\prime}}$) represent the bosonic annihilation (creation) operators corresponding to the energy mode $\omega^{\prime}$. These operators, having the unit of $1/\sqrt{\omega^{\prime}}$, satisfy the commutation relation $[b_{\omega^{\prime}}^{j},b_{\omega^{\prime\prime}}^{j\dagger}]=\delta(\omega^{\prime}-\omega^{\prime\prime})$. Here, $\tilde{\omega}$ is an arbitrary constant with units of frequency, and $\Omega_j$ denotes the cutoff frequency for the $j^{\text{th}}$ bath. We take the coupling between the $j^{\text{th}}$ qubit and $j^{\text{th}}$ bath to be
\begin{equation}
    H_{SB_j}=\hbar \sqrt{\tilde{\omega}} \int_{0}^{\Omega_j} d\omega^{\prime} \chi_j(\omega^{\prime}) \Big( \sigma^j_{+} b^j_{\omega^{\prime}} + \sigma^j_{-} b^{j\dagger}_{\omega^{\prime}}  \Big),
    \label{eq:bath_int}
\end{equation}
where $\chi_j(\omega^{\prime})$, a dimensionless function of $\omega^{\prime}$, modulates the coupling strength of the local interaction between the $j^{\text{th}}$ qubit and $j^{\text{th}}$ bath. Specifically, we have $\tilde{\omega}\chi_j^2(\omega^{\prime}) = J_j(\omega^{\prime})$, where $J_j(\omega^{\prime})$ denotes the spectral density function of the $j^{\text{th}}$ bosonic bath. For our purposes, we adopt the Ohmic spectral density function, which takes the explicit form $J_j(\omega^{\prime}) = \alpha_j \omega^{\prime} \exp(-\omega^{\prime}/\Omega_j)$. Here $\alpha_j$s are dimensionless parameters. For the choice of the system-bath intection Hamiltonian given in Eq.~(\ref{eq:bath_int}), we obtain the decay constants as
\begin{eqnarray}
\label{gamma}
\gamma_{j}(\omega) &=& J_j(\omega) [1+f(\omega,T_j)] \quad \quad  \omega>0 \nonumber \\
                  &=& J_j(|\omega|) f(|\omega|,T_j) \quad \quad \quad \;  \omega<0,
\end{eqnarray}
where $f(\omega,T_j)=[\exp(\hbar \omega/k_BT_j)-1]^{-1}$ being the Bose-Einstein distribution function. Suppose we have two eigenvectors, $\ket{k}$ and $\ket{l}$, of the system Hamiltonian $\mathcal{H}_0+\mathcal{H}^1_{\text{int}}$, corresponding to eigenvalues $\lambda_k$ and $\lambda_l$ respectively. The Lindblad operators for the interaction Hamiltonian $H_{SB_j}$ [Eq.~(\ref{eq:bath_int})] can then be written as
\begin{equation}
    A_j(\omega)=\sum_{\omega=\frac{1}{\hbar}(\lambda_l-\lambda_k)}\ket{k}\bra{k}\sigma^j_{x}\ket{l}\bra{l}.
\end{equation}

\section{Performance of QHT with Identical Bath Temperatures}
\label{Sec:same_temp}

In our QHT model, we have considered thermal baths at different temperatures. One may now naturally ask whether the transformation phenomenon can still occur if all the baths are maintained at the same temperature or if only a single thermal bath is used. To explore this, we investigate the single-bath scenario and illustrate the corresponding step-up and step-down behavior in Fig.~\ref{Fig:same_temp}. As seen from the left (step-down) and right (step-up) panels, the resulting temperature gradients and their differences across the primary and secondary junctions are significantly smaller compared to the results in Fig.~\ref{Fig:H_int_111-000} and Fig.~\ref{Fig:step-up_3q} of the main text, where all parameters except the bath temperatures are kept the same. Notably, we find that step-up transformer behavior cannot be achieved in the steady-state regime, as shown in the right panel of Fig.~\ref{Fig:same_temp}. We also examine a range of bath temperatures, but the outcomes remain qualitatively similar.

The underlying reason is as follows. When all three baths are at the same temperature, and the self-contained condition $\mathcal{E}_3=-\mathcal{E}_1-\mathcal{E}_2$ is employed, the probability of the system initially being in the states $\ket{000}$ and $\ket{111}$ becomes equal, i.e.,
$(1-p_1^{\text{in}})(1-p_2^{\text{in}})(1-p_3^{\text{in}}) = p_1^{\text{in}} p_2^{\text{in}} p_3^{\text{in}}$. As a result, activating $\mathcal{H}_{\text{int}}^1$ leads to equal likelihood of transition between these two states in either direction, which reduces the net directional flow and significantly diminishes the temperature gradients. For this reason, we chose to focus our analysis on configurations with distinct bath temperatures, which allow the machine to operate more effectively in both step-up and step-down regimes.

\begin{figure*}[htb]
	\includegraphics[scale=0.32]{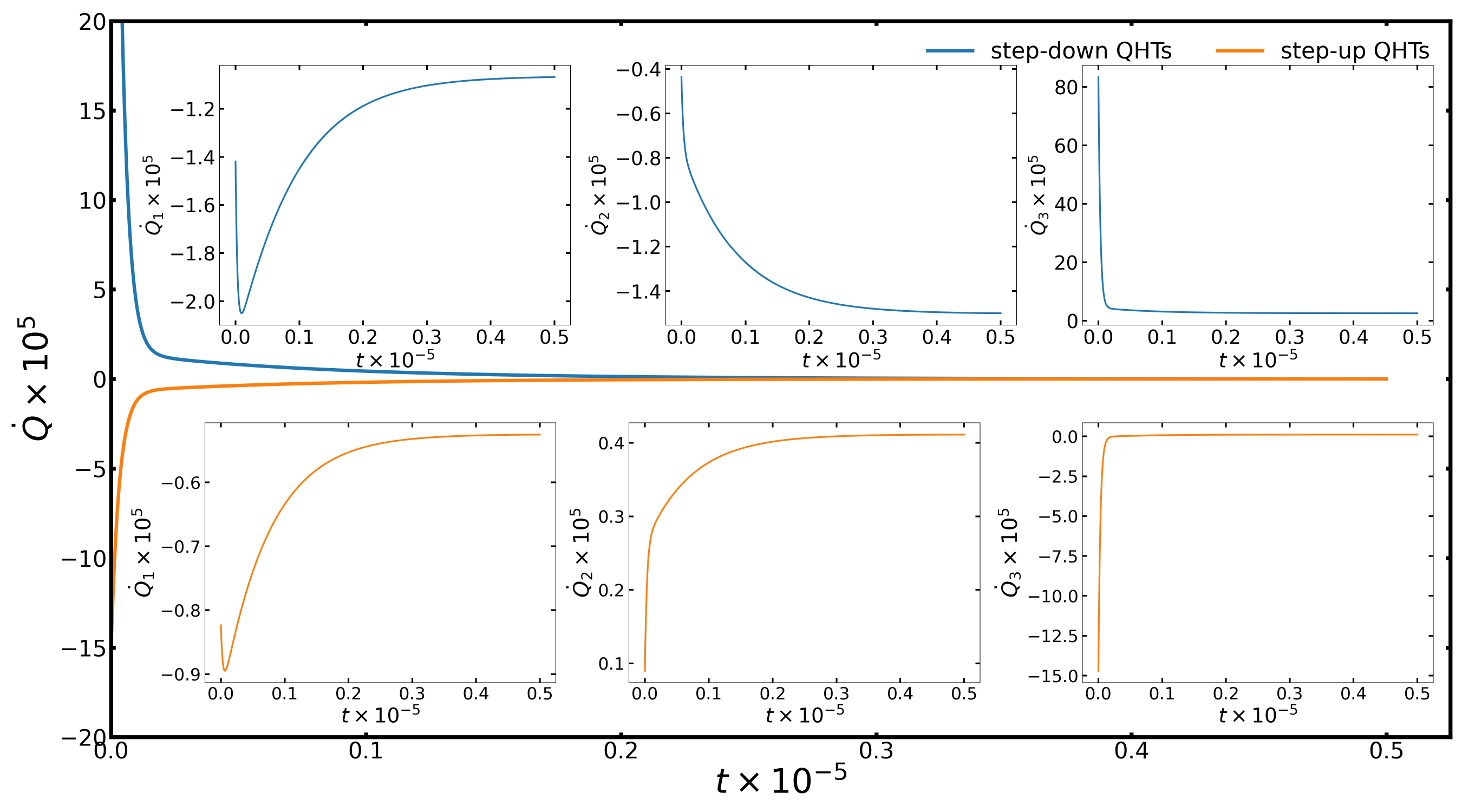}
		\caption{Time evolution of the heat currents for the step-down and step-up modes corresponding to the interaction Hamiltonian $\mathcal{H}_{\text{int}}^1$. The main figure shows the total heat current $\dot{Q}(t)$ between the system and the baths. The insets display the local heat currents $\dot{Q}_1(t)$, $\dot{Q}_2(t)$, and $\dot{Q}_3(t)$ (blue for step-down, orange for step-up). Here, $\tau_3=3.0$, and all the other parameters are same as in Fig.~\ref{Fig:H_int_111-000} for the step-down mode and Fig.~\ref{Fig:step-up_3q} for the step-up mode.   All quantities plotted here are dimensionless.}

		\label{fig:Q-diagram}
	\end{figure*}

\section{Behavior of the heat currents}
\label{app:heat_current}


In Fig.~\ref{fig:Q-diagram}, we present the behavior of the total heat current $\dot{Q}(t)$ as a function of time for the parameter set of Fig.~\ref{Fig:H_int_111-000} with, $\tau_3=3.0$. 
For the step-down mode of QHT, $\dot{Q}(t)$ begins with a positive value, indicating that the system initially absorbs energy from the baths. As the system evolves toward its steady state, the heat current monotonically decays to zero, which means there is no net energy exchange between the system and the environments. For the step-up mode, the behavior is analogous, except that $\dot{Q}(t)$ starts with a negative value, reflecting initial energy loss from the system to the baths. The insets of Fig.~\ref{fig:Q-diagram} show the local heat currents $\dot{Q}_1(t), \dot{Q}_2(t), \dot{Q}_3(t)$ for both modes. Local heat currents are defined as $\dot{Q}_j(t)=\frac{\hbar}{\mathcal{K}}\Tr[\mathscr{H} \mathcal{L}_j(\rho(t))]$, where 
$\mathcal{L}_j(\rho(t))$ is the dissipator acting on the $j$th qubit. Importantly, even at the steady state --- where the total heat current satisfies $\dot{Q}_1+\dot{Q}_2+\dot{Q}_3=0$, each individual $\dot{Q}_j$ remains non-zero. 

We observe that, when $\dot{Q}_j(t)<0$, the temperature of the $j$th qubit, $T_j(t)$, increases over time, and when $\dot{Q}_j(t)>0$, $T_j(t)$ decreases. Therefore, the heat currents clearly reveal whether a given qubit is being heated or cooled. 

\section{Total heat flow from a junction}
\label{app:tot_q}

If we now consider the total heat absorbed or released from the junctions up to a certain time instead of the heat currents, we observe a behavior similar to that presented in Table \ref{Ref3:Tab:Q-dot} of the main text.
We define the total heat exchanged between the system and the $j$‑th bath over the evolution time $t_f$ as
\[
Q_j^{ \text{tot}} =  \int_{0}^{t_f} \dot{Q}_j(t)\, dt.
\]
Here, $Q_j^{\text{tot}} > 0$ corresponds to net heat absorbed by the qubit from bath $j$, and $Q_j^{\text{tot}} < 0$ corresponds to net heat released to the bath by the qubit.

\begin{table}[h]
\centering
\begin{tabular}{c|c|c|c}
$\tau_3$ & $Q_{ p}^{ \text{tot}}$ & $Q_{ s}^{ \text{tot}}$ & Analysis \\
\hline
\multicolumn{4}{c}{\textbf{In step-down mode Conditions}} \\
\hline
3.0 & -0.972 & 0.832 & Primary loss $>$ Secondary gain \\
2.5 & -0.745 &  0.637 & Primary loss $>$ Secondary gain \\
2.0 & -0.438 & 0.374 &Primary loss $>$ Secondary gain \\
\hline
\multicolumn{4}{c}{\textbf{In step-up mode Conditions}} \\
\hline
3.0 & -0.080 & 0.124 & Primary loss  $<$ Secondary gain \\
2.5 & -0.056 & 0.112  & Primary loss  $<$ Secondary gain \\
2.0 & -0.023 & 0.095  & Primary loss  $<$  Secondary gain \\
\end{tabular}
\caption{Total heat gained or lost by the primary and secondary junctions for different values of $\tau_3$ in the step-down and step-up modes. All other parameters are fixed as in Fig.~\ref{Fig:H_int_111-000} and Fig.~\ref{Fig:step-up_3q} of the main text for the step-down and step-up modes, respectively. Here, $t_f=0.5\times 10^5$. The ``Analysis" column compares the relative magnitudes of heat gain and loss.
All the quantities whose numerical values are presented here are dimensionless.
}
\label{Tab:1}
\end{table}

We define the total heat absorbed or released by the primary and secondary junctions as $Q_{p}^{\text{tot}}$  = $Q^{\text{tot}}_1 + Q^{\text{tot}}_2$ and $Q_{s}^{\text{tot}}$ = $Q^{\text{tot}}_2 + Q^{\text{tot}}_3$, respectively,
where a positive value denotes net heat absorbed by the junction, and a negative value denotes net heat released by the junction. For the step-up mode (Fig.~\ref{Fig:step-up_3q} of the main text with $\tau_3 =3.0$ and $t_f=0.5\times 10^5$), we obtain
\[
Q_{p}^{\text{tot}} = -0.080, 
\quad 
Q_{s}^{\text{tot}} = 0.124.
\]
Hence, the primary junction releases heat, while the secondary junction absorbs an amount of heat that exceeds the heat released by the primary.
For the step-down mode (Fig.~\ref{Fig:H_int_111-000} with $\tau_3=3.0$ and $t_f=0.5\times 10^5$), we obtain
\[
Q_{p}^{\text{tot}} = -0.972, 
\quad 
Q_{s}^{\text{tot}} = 0.832.
\]
In this case, the primary junction loses more heat than the secondary junction gains. To examine the generality of this behavior, we vary the temperature of the third bath, $\tau_3$, and obtain the results summarized in Table~\ref{Tab:1}. All other parameters are kept fixed as in Fig.~\ref{Fig:H_int_111-000} and Fig.~\ref{Fig:step-up_3q} of the main text for the step-down and step-up modes, respectively.


\begin{figure*}[htb]
	\includegraphics[scale=0.5]{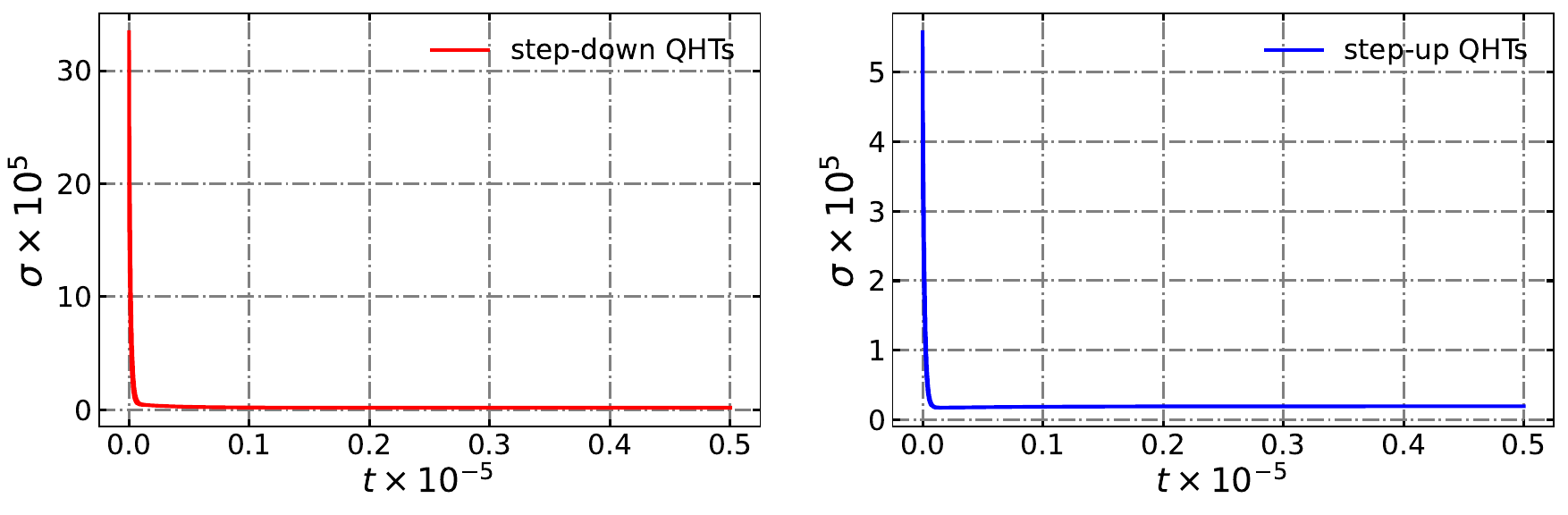}
		\caption{Time evolution of the entropy production rate  for the interaction Hamiltonian $\mathcal{H}_{\text{int}}^1$. Here, $\sigma(t)$ is plotted for both step-down and step-up modes with $\tau_3=3.0$. All the other parameters are the same as in Fig.~\ref{Fig:H_int_111-000} for the step-down mode and Fig.~\ref{Fig:step-up_3q} for the step-up mode. All quantities plotted here are dimensionless.}
        \label{Fig:sigma}
	\end{figure*}

From Table \ref{Tab:1}, we observe that for all cases discussed in the manuscript, within the three-qubit scenario and for the interaction Hamiltonian $\mathcal{H}_{\text{int}}^1$, this behavior remains consistent across different values of $\tau_3$. 

\section{Validation of the first and second laws of thermodynamics }
\label{app:laws}

The first law of thermodynamics for the open quantum evolution discussed in the main text reads
\begin{equation}
    \dot{Q}(t) \;=\; \frac{\hbar}{\mathcal{K}}\frac{\mathrm{d}}{\mathrm{d}t}\big\langle \mathscr{H}_0 + \mathscr{H}_{\mathrm{int}}^{x}\big\rangle=\frac{\hbar}{\mathcal{K}}\frac{\mathrm{d}}{\mathrm{d}t}\Tr\big[\mathscr{H}\rho(t)\big].
\end{equation}
We have incorporated the $\frac{\hbar}{\mathcal{K}}$ term to make $\dot{Q}(t)$ dimensionless. As the Hamiltonian $\mathscr{H}=\mathscr{H}_0+\mathscr{H}_{\text{int}}$ is time independent, we obtain
\begin{equation}
    \dot{Q}(t) \;=\frac{\hbar}{\mathcal{K}}\Tr\big [\mathscr{H}\frac{\mathrm{d\rho(t)}}{\mathrm{d}t}\big ]=\frac{\hbar}{\mathcal{K}}\Tr\big[\mathscr{H}\mathcal{L}(\rho(t)) \big ].
\end{equation}
Thus, the expression we use for $\dot{Q}(t)$ is fully consistent with the first law of thermodynamics.

The quantum formulation of the second law of thermodynamics for the Markovian evolution is expressed through Spohn’s inequality,
\begin{equation}
    \sigma(t)=\dot{S}(t)-\sum_j\frac{\dot{Q}_j(t)}{T_j}\ge 0,
\end{equation}
where $S(t)=-\Tr[\rho(t)\ln \rho(t)]$ is the von Neumann entropy of the system and $\sigma(t)$ is the entropy production rate. This relation holds for any Markovian dynamics. In Fig.~\ref{Fig:sigma}, we show $\sigma(t)$ for the case depicted in Fig.~\ref{Fig:H_int_111-000} (step-down mode) and Fig.~\ref{Fig:step-up_3q} (step-up mode), both with $\tau_3=3.0$. For both step-down and step-up modes, $\sigma \ge 0$, demonstrating the validity of Spohn’s theorem.

\bibliography{QHT}
\end{document}